\documentclass[prd,aps,floats,floatfix,superscriptaddress,preprintnumbers,showpacs,eqsecnum,nofootinbib,twocolumn]{revtex4}
\usepackage{amsfonts,amsmath,epsfig,amssymb,latexsym,color,graphicx,array,subfigure,bm,float }
\usepackage{accents}

\definecolor{red}{rgb}{1,0,0}
\definecolor{blue }{rgb}{0,0,1}
\definecolor{green}{rgb}{0,1,0}

\newcommand{\mybar}[1]{\accentset{\rule{0.8em}{0.4pt}}{#1}}

\newcommand{\bea}{\begin{eqnarray}}
\newcommand{\ena}{\end{eqnarray}}
\newcommand{\beann}{\begin{eqnarray*}}
\newcommand{\enann}{\end{eqnarray*}}

\newcommand{\dsl}{\pa \kern-0.5em /}

\newcommand{\pa}{\partial}

\newcommand{\nn}{\nonumber\\}

\newcommand{\vect}[1]{\!\!\mbox{ \boldmath $#1$}}

\newcommand{\Mbar}{\mybar{M}}

\begin{document}

\title{Self-gravitating Schr\"odinger–Poisson Solitons in Curved Spacetime}

\author{Kei-ichi Maeda}
\affiliation{
Department of Physics, Waseda University, 3-4-1 Okubo, Shinjuku, Tokyo 169-8555, Japan}
\affiliation{Center for Gravitational Physics and Quantum Information, Yukawa Institute for Theoretical Physics, Kyoto University, 606-8502, Kyoto, Japan}

\begin{abstract}

We formulate a description of self-gravitating Schr\"odinger–Poisson (SP) solitons in curved spacetime using a local inertial frame centered on the soliton. In the weak-field and non-relativistic regime, the effect of the background spacetime is incorporated into the SP equations through an effective tidal potential determined by the spacetime curvature. We apply this framework to a soliton moving along a circular equatorial geodesic around a Kerr black hole. The tidal field contains stationary and time-dependent quadrupolar components, which induce a deformation of the soliton and an associated mass quadrupole moment. We formulate the quadrupolar response in terms of the tidal deformability and the dimensionless Love number. We then estimate the gravitational radiation associated with the rotating deformed soliton and evaluate the corresponding characteristic strain for representative soliton sizes and orbital radii. A comparison with representative sensitivity curves of planned space-based gravitational-wave detectors identifies parameter regimes in which the estimated characteristic strain may lie within their sensitivity ranges. Our framework provides a systematic basis for investigating the tidal response of self-gravitating bosonic solitons in black-hole spacetimes and their potential gravitational-wave signatures.
\end{abstract}

\maketitle

\section{Introduction}

Dark matter is one of the outstanding problems in modern cosmology. Its existence is supported by a wide range of astrophysical and cosmological observations, while its microscopic nature remains unknown. A variety of candidates have been proposed, including new elementary particles \cite{BertoneHooperSilk2005} and primordial black holes \cite{CarrKuhnel2020}. Another possibility is that dark matter is composed of bosonic fields that form self-gravitating localized configurations, which have attracted considerable attention in relativistic astrophysics. Such objects include boson stars \cite{Kaup1968,RuffiniBonazzola1969,LieblingPalenzuela2023}, axion stars \cite{BraatenZhang2019}, and oscillatons \cite{Page2004}, and have been studied both as possible compact objects and as possible realizations of dark matter. Boson stars were introduced in the pioneering works of Kaup and of Ruffini and Bonazzola \cite{Kaup1968,RuffiniBonazzola1969}, and their equilibrium and dynamical properties have subsequently been studied extensively \cite{LieblingPalenzuela2023}. Scalar-field configurations have also been considered in the context of ultralight dark matter, where the wave nature of the field naturally leads to self-gravitating solitonic structures \cite{HuiOstrikerTremaineWitten2017,Hui2021}. Related non-relativistic descriptions have been developed for axion stars and other scalar-field configurations \cite{BraatenMohapatraZhang2016NR,BraatenZhang2019,Aoki:2017ixz}.

In the weak-field and non-relativistic regime, the Einstein-Klein-Gordon system reduces to the Schr\"odinger-Poisson (SP) equations. The SP system provides an effective description of self-gravitating bosonic condensates and has consequently been widely used in studies of ultralight dark matter and bosonic compact objects \cite{HuiOstrikerTremaineWitten2017,Hui2021}. Recent studies have also investigated the interplay between black holes and ultralight dark matter, including self-gravitating solitonic or boson-star configurations in the vicinity of black holes \cite{BucciottiTrincherini2023}. The possible gravitational-wave signatures of dark-matter environments containing solitonic cores have further been explored \cite{KadotaKimKoYang2024}.

The response of compact bosonic objects to an external gravitational field provides another important aspect of their physics. 
Tidal deformation can probe the internal structure of a compact object, and the induced quadrupole moment is commonly characterized by the tidal deformability or Love number \cite{Hinderer2008,BinningtonPoisson2009,Damour_2009}. Tidal Love numbers have been studied for boson stars and other exotic compact objects, together with their possible effects on gravitational-wave observations \cite{FlanaganHinderer2008,CardosoFranzinMaselliPaniRaposo2017,SennettHindererSteinhoffBuonannoOssokine2017}. More recently, the tidal response and Love numbers of axion stars have been investigated using relativistic tidal perturbation theory \cite{ChenHuangZhaoWang2024}.
 These studies show that the tidal response of scalar-field configurations can provide information about their internal structure.

The description of self-gravitating systems in an external gravitational field using a local inertial frame has also been developed in several contexts. Fermi normal coordinates provide a natural local description of the spacetime geometry in the vicinity of a freely falling worldline \cite{Manasse1963,Gorbatsievich_Bobrik}. In particular, Ishii, Shibata, and Mino studied the tidal deformation and disruption of a Newtonian self-gravitating star orbiting a Kerr black hole using Fermi normal coordinates \cite{IshiiShibataMino2005}.
We have also developed a local-inertial frame formulation for self-gravitating binary systems in the vicinity of black holes \cite{Maeda:2023tao,Maeda:2023uyx,Maeda:2025row,Maeda:2026}. For binary systems on circular orbits, 
we constructed local inertial reference frames based on Fermi–Walker transport and formulated the Newtonian dynamics of the binary in the tidal field of a Schwarzschild black hole and subsequently of a Kerr black hole \cite{Maeda:2023tao,Maeda:2023uyx}. This formulation was then extended to more general orbital configurations, including spherical and eccentric orbits \cite{Maeda:2025row,Maeda:2026}. 
 These studies provide a systematic framework for describing the dynamics of a self-gravitating system.

The above developments naturally raise the question of how a similar local-inertial frame description can be formulated for a self-gravitating scalar-field configuration. In particular, it is important to clarify how the tidal field of a black hole enters the dynamics of a Schr\"odinger--Poisson soliton and how the resulting tidal deformation may be related to gravitational-wave emission. 

In this work, we formulate a curved-spacetime extension of the Schr\"odinger-Poisson description for a self-gravitating soliton. We employ a local inertial frame centered on the soliton and describe the external gravitational field through the tidal field of the background spacetime. We formulate the tidal deformation and the induced quadrupole moment of the soliton and discuss the gravitational-wave emission associated with the rotating soliton.

The paper is organized as follows. In Sec.~II, we formulate the Schr\"odinger--Poisson system for a self-gravitating soliton in curved spacetime using a local inertial frame. In Sec.~III, we derive the tidal field in Kerr spacetime and formulate the tidal deformation of the soliton. In Sec.~IV, we discuss a stationary rotating soliton. In Sec.~V, we define the tidal deformability and Love number. In Sec.~VI, we estimate the gravitational-wave signal and compare its characteristic strain with representative detector sensitivities. Finally, Sec.~VII summarizes our results.

\section{Soliton in curved spacetime}
Suppose that spacetime is curved, with the
  metric given by
\bea
d\bar s^2=\bar g_{\mu\nu}dx^\mu dx^\nu
\label{background_metric}
\,.
\ena
In this background spacetime, 
we consider a self-gravitating boson condensate consisting of a complex massive scalar field $\Psi$, 
whose action is given by
\bea
S=-\int d^4x \sqrt{-g}\left[\nabla_\mu \Psi^* \nabla^\mu \Psi +m_S^2\Psi^*\Psi\right]
\ena
where  $m_S$ is the mass of the scalar field. 
We call this condensate a soliton.
The corresponding Klein-Gordon equation is
\begin{equation}
(\Box_{\bar g}-m_S^2)\Psi=0 .
\end{equation}

To describe a soliton embedded in a background
gravitational field (\ref{background_metric}), we introduce a local inertial frame, 
with coordinates $(x^{\hat \mu})=(\tau, \hat{\vect{x}})$. 
We assume that the size of a soliton is much smaller than the curvature radius of 
a background spacetime $\ell_{\bar{\cal R}}$.
We also assume that a condensate is described by Newtonian dynamics in this local inertial frame. 
We introduce the non-relativistic decomposition of the scalar field,
\begin{equation}
\Psi(\tau, \hat{\vect{x}})
=
\frac{1}{\sqrt{2m_S}}
e^{-im_S\tau }\psi(\tau, \hat{\vect{x}}),
\end{equation}
where the slowly varying wave function $\psi$ satisfies
\begin{equation}
|\partial_\tau \psi|, |\partial_{\hat j}\psi| \ll m_S|\psi|.
\end{equation}

Assuming the metric in the local inertial frame is given by $g_{\hat \mu \hat \nu}=
\eta_{\hat \mu \hat \nu}+h_{\hat \mu \hat \nu}$, and 
keeping the leading-order terms in the non-relativistic expansion, the
Klein-Gordon equation reduces to a Schr\"odinger equation with a Newtonian potential $\Phi=-h_{\hat 0\hat 0}/2$,
\begin{equation}
i\partial_\tau\psi
=
-\frac{1}{2m_S}\hat \nabla^2\psi
+
m_S\Phi\psi 
\,.
\label{Schrodinger_eq}
\end{equation}

Suppose the center of mass of a soliton follows an observer's orbit.
In the vicinity of the
observer, the metric expansion is given by 
Eq.(\ref{metric_reference_frame}) (see Appendix \ref{proper reference frame}).
The Newtonian potential caused by the external force field is obtained by the 
perturbed metric $\epsilon_{\hat 0\hat 0}$ in the 
local inertial frame.  
There are two contributions; one is the tidal force given by the Riemann curvature
 and the other is the 
inertial force which consists of the acceleration and rotation.
We then describe them as
\bea
\Phi_{\rm tidal}&=&\frac{1}{2}
\bar{\cal R}_{\hat 0 \hat k \hat 0 \hat \ell }
x^{\hat k}  x^{\hat \ell}  ,
\label{tidal_potential0}
\\
\Phi_{\rm inertial}&=&\frac{1}{2}\left[2a_{\hat k}
x^{\hat k}
-
\omega_{\hat j\hat k} \omega^{\hat j}_{~\hat \ell} 
x^{\hat k}  x^{\hat \ell} +\left(a_{\hat k} x^{\hat k}\right)^2   \right]
\,,~~~
\label{inertial_potential0}
\ena
where $\bar{\cal R}_{\hat \mu \hat \nu \hat \rho \hat \sigma }$ is the Riemann curvature of the background metric (\ref{background_metric}), and $a_{\hat k}$ and $\omega^{\hat j}_{~\hat \ell}$ are 
the acceleration and angular velocity of the observer.

When we consider a self-gravitating soliton, we also have to include the Newtonian potential 
$\Phi_{\rm self}$ given by
the scalar field itself, which satisfies the Poisson equation
\bea
\hat \nabla^2\Phi_{\rm self}=4\pi G
\rho_S
\label{Poisson_eq}
\,,
\ena
where $\rho_S\equiv m_S|\psi|^2$ is the mass density of a soliton in Newtonian approximation.

We finally obtain  the Schr\"odinger-Poisson  (SP) system in the curved-spacetime, as
Eq. (\ref{Schrodinger_eq}) with 
\begin{equation}
\Phi\equiv \Phi_{\rm self}
+\Phi_{\rm inertial}
+\Phi_{\rm tidal}
\,,
\end{equation}
and Eq. (\ref{Poisson_eq}).
This set of equations provides an effective description of self-gravitating
bosonic objects subject to an external field.

To solve these equations, we  
have to fix the observer's orbit, that is, the origin of the 
local inertial frame.
One of the most convenient choices is to require that 
the center of mass of the soliton coincides with the origin $\hat{\vect{x}}=0$.

The center of mass is defined by
\bea
\vect{R}_{\rm CM}(\tau)\equiv {1\over M_{\rm soliton}}\int d^3\hat{\vect{x}} \, \rho_S(\tau,\hat{\vect{x}} )  \, \hat{\vect{x}}
\,, 
\ena
where
\bea
M_{\rm soliton}&=&
\int d^3\hat{\vect{x}} \, \rho _S
\ena
is the mass of a soliton.

To satisfy the condition of $\vect{R}_{\rm CM}(\tau)=0$ at all times, we have to impose
\bea
\dot{\vect{R}}_{\rm CM}(\tau)=0\,,~\ddot{\vect{R}}_{\rm CM}(\tau)=0
\,,
\ena
where a dot denotes the derivative with respect to the proper time $\tau$. 

Using the Sch\"odinger equation (\ref{Schrodinger_eq}), we find
\bea
\dot{R}_{\rm CM}^{\hat k}
&=&{m_S\over M_{\rm soliton}}\int d^3\vect{x} J^{\hat k}
\ena
where
the flux density $\vect{J}$ is defined by
\bea
\vect{J}\equiv -{i\over 2m_S}\left[\psi^*\hat{\vect{\nabla}} \psi-(\hat{\vect{\nabla}}\psi^*)\psi\right]
\,.
\ena
We also find
\bea
\ddot{R}_{\rm CM}^{\hat k}&=&{m_S\over M_{\rm soliton}}\int d^3\hat{\vect{x}} \dot{J}^{\hat k} 
\nn
&=&
-{1\over M_{\rm soliton}}\int d^3 \hat{\vect{x}}  \,(\hat{\nabla}^{\hat k} \Phi) \rho_S
\ena

Since $\Phi_{\rm self}$ is a self-gravitating potential, using the Poisson equation (\ref{Poisson_eq}), 
we find 
\beann
\int d^3 \hat{\vect{x}} \,(\nabla^{\hat k} \Phi_{\rm self}) \rho_S=0
\,.
\enann
The other two potentials, $\Phi_{\rm tidal}$ and $\Phi_{\rm inertial}$ contribute to $\ddot{R}_{\rm CM}^k$.
From Eqs. (\ref{tidal_potential0}) and (\ref{inertial_potential0}), we find 
 $\Phi'\equiv\Phi_{\rm tidal}+\Phi_{\rm inertial}$ evaluated in the local inertial frame
are quadratic functions of  $x^{\hat k}$, i.e.,
\beann
\Phi'(\vect{x},\tau)=\Phi'_0+(\hat{\nabla}_{\hat k} \Phi')_0 x^{\hat k} +{1\over 2}(\hat{\nabla}_{\hat k} \hat{\nabla}_{\hat \ell} \Phi')_0 x^{\hat k} x^{\hat \ell}
\enann
where 
\beann
\Phi'_0&\equiv& \Phi'(\tau,\hat{\vect{0}})\,,
\nn
 (\hat{\nabla}_{\hat k} \Phi')_0&\equiv&
 (\hat{\nabla}_{\hat k} \Phi')(\tau,\hat{\vect{0}})\,, 
 \nn
 (\hat{\nabla}_{\hat k} \hat{\nabla}_{\hat \ell} \Phi')_0&\equiv&
  (\hat{\nabla}_{\hat k} \hat{\nabla}_{\hat \ell} \Phi')(\tau,\hat{\vect{0}}).
 \enann
We then obtain
\beann
\ddot{R}_{\rm CM}^{\hat k}&=&-(\hat{\nabla}^{\hat k}  \Phi')_0-(\hat{\nabla}^{\hat k} \hat{\nabla}_{\hat \ell }
\Phi')_0 R_{\rm CM}^{\hat \ell}(\tau)
\enann
Assuming $\vect{R}_{\rm CM}=0$ and $\dot{\vect{R}}_{\rm CM}=0$ initially, 
we find $\vect{R}_{\rm CM}=0$ anytime if 
\bea
(\hat{\nabla}^{\hat k} \Phi')_0=0
\,.
\ena
We then find the condition is 
\bea
a^{\hat k}=0 
\ena
because only the term with $a_{\hat k}$ in $\Phi_{\rm inertial}$ is a linear function of $x^{\hat k}$.
If we consider higher-order terms of $x^{\hat k}$ such as a cubic term, 
we find non-zero acceleration $a^{\hat k}$ to set up the center-of-mass at the origin.

In the present approximation, 
we find the  SP equations in the non-rotating ($\omega_{\hat j\hat k}=0$) inertial frame, 
\begin{equation}
i\dot{\psi}
=
\left[
-\frac{\hat \nabla^2}{2m_S}
+
m_S(\Phi_{\rm self}
+\Phi_{\rm tidal})
\right]\psi ,
\label{Schrodinger}
\end{equation}
together with
\begin{equation}
\hat \nabla^2\Phi_{\rm self}
=
4\pi Gm_S|\psi|^2 .
\label{Poisson}
\end{equation}
The center of mass of a soliton is the origin in this coordinate system 
and follows a geodesic in a given background spacetime
(\ref{background_metric}).

\begin{widetext}

\section{Soliton in Kerr spacetime}

In what follows, we consider a self-gravitating soliton orbiting around a Kerr black hole, 
whose spacetime metric in 
Boyer--Lindquist coordinates  $(\bar t, \bar r, \bar \theta, \bar \phi)$ is written as
\bea
d\bar s^2
&=&
-\frac{\Delta}{\Sigma}\left[d\bar{t}-a\sin^2\bar \theta d\bar{\phi}\right]^2
+\frac{\sin^2\bar{\theta}}{\Sigma}\left[
(\bar{r}^2+a^2)d\bar{\phi}-ad\bar{t}\right]^2+\frac{\Sigma}{\Delta}d\bar{r}^2+\Sigma d\bar{\theta}^2
\,,
\label{Kerr_metric}
\ena
\end{widetext}
where
\beann
\Delta&=& \bar{r}^2-2\mybar{M}\bar{r}+a^2
\\
\Sigma&=& \bar{r}^2+a^2 \cos^2\bar{\theta}
\,.
\enann
with 
 $\mybar{M}\equiv GM$.
 $M$ and $a$ are the mass and the specific angular momentum
 of a black hole, respectively.
 
 To describe a soliton in Kerr spacetime, we first determine 
the tidal potential $\Phi_{\rm tidal}$, i.e., 
 the Riemann curvature components in a non-rotating local inertial frame, which 
coordinates are described by $(x^{\tilde \mu})= (\tau, \mathsf{x},\mathsf{y},\mathsf{z})$.
 In this paper, we consider only a soliton moving on a circular orbit with the radius $\bar{r}_0$.

\subsection{Tidal potential in Kerr spacetime}
We consider the center of mass of a soliton moving on a circular orbit of radius $\bar r_0$ in the 
equatorial plane in the Kerr spacetime.
As we summarize in Appendix \ref{Riemann_curvature},  
the relevant components  of the Riemann curvature 
 in the non-rotating inertial frame, whose indices are denoted by tildes, are given by
 \beann
 \bar{\cal R}_{\tilde 0\tilde 1\tilde 0\tilde 1} &=&\bar{\cal R}_{\hat 0 \hat 1\hat 0\hat 1}\cos^2\varphi 
 +\bar{\cal R}_{\hat 0 \hat 2\hat 0\hat 2}\sin^2\varphi
 \label{R0101}
 \\
\bar{\cal R}_{\tilde 0\tilde 2\tilde 0\tilde 2} &=&\bar{\cal R}_{\hat 0 \hat 1\hat 0\hat 1}\sin^2\varphi
+\bar{\cal R}_{\hat 0 \hat 2\hat 0\hat 2}\cos^2\varphi
 \label{R0202} \\
\bar{\cal R}_{\tilde 0\tilde 3\tilde 0\tilde 3} &=&\bar{\cal R}_{\hat 0 \hat 3\hat 0\hat 3}
 \label{R0303} \\
\bar{\cal R}_{\tilde 0\tilde 1\tilde 0\tilde 2} &=&\left(
\bar{\cal R}_{\hat 0 \hat 1\hat 0\hat 1}-\bar{\cal R}_{\hat 0 \hat 2\hat 0\hat 2}\right)\cos\varphi\sin\varphi
 \label{R0102}
\nonumber \,,
 \enann
 where
 \bea
 \varphi\equiv \sigma \mathfrak{w}_{\rm R} \tau\,,~~\mathfrak{w}_{\rm R}\equiv\frac{\mybar{M}^{1/2}}{\bar{r}_0^{3/2}}
 \ena
are the rotation angle and angular velocity in the $xy$ plane measured relative to the rotating inertial frame, whose indices are denoted by hats. 
 Here, $\sigma=\pm 1$ corresponds to prograde and retrograde 
 motion, respectively.

The components of the Riemann curvature $\bar{\cal R}_{\hat 0 \hat i\hat 0\hat j}$ in the rotating inertial frame, with the coordinates $(x^{\hat \mu})= (\tau, x,y,z)$,
 are given by
\beann
&&
\bar{\cal R}_{\hat 0 \hat 1\hat 0\hat 1}
=-{\mybar{M}\over   \bar{r}_0^3}{G^2_\sigma(\bar{r}_0) \over F^2_\sigma(\bar{r}_0)}
\,,~
\\
&&
\bar{\cal R}_{\hat 0 \hat 2\hat 0\hat 2}={\mybar{M}\over \bar{r}_0^3}
\,,~
\\
&&
\bar{\cal R}_{\hat 0 \hat 3\hat 0\hat 3}=
{\mybar{M}\over  \bar{r}_0^3}{H^2_\sigma(\bar{r}_0) \over F^2_\sigma(\bar{r}_0) }
\,,
\enann
where
\beann
F^2_\sigma(\bar{r}_0) &\equiv&\Delta(\bar{r}_0)-S_\sigma(\bar{r}_0)
\\
G^2_\sigma(\bar{r}_0) &\equiv&2\Delta(\bar{r}_0)+S_\sigma(\bar{r}_0)
\\
H^2_\sigma(\bar{r}_0) &\equiv&\Delta(\bar{r}_0)+2S_\sigma(\bar{r}_0)
\,.
\enann
with
\beann
S_\sigma(\bar{r}_0)\equiv \mybar{M}\bar{r}_0+a^2-2a\sigma\sqrt{\mybar{M}\bar{r}_0}
\,.
\enann
 \begin{widetext}
We then find the tidal potential as
\bea
\Phi_{\rm tidal}
&=&
-\frac{1}{2}\left[ \bar{\cal R}_{\tilde 0\tilde 1\tilde 0\tilde 1} \mathsf{x}^2+ \bar{\cal R}_{\tilde 0\tilde 2\tilde 0\tilde 2}\mathsf{y}^2+
 \bar{\cal R}_{\tilde 0\tilde 3\tilde 0\tilde 3} \mathsf{z}^2+2 \bar{\cal R}_{\tilde 0\tilde 1\tilde 0\tilde 2}\mathsf{x}\mathsf{y} \right]
 \nn
 &=&
 -\frac{1}{2}\left[ \bar{\cal R}_{\hat 0\hat 1\hat 0\hat 1}\left(\mathsf{x}\cos\varphi+\mathsf{y}\sin\varphi\right)^2
 + \bar{\cal R}_{\hat 0\hat 2\hat 0\hat 2} \left(\mathsf{x}\sin\varphi-\mathsf{y}\cos\varphi\right)^2+
 \bar{\cal R}_{\hat 0\hat 3\hat 0\hat 3} \mathsf{z}^2 \right]
  \nn
 &=&
 -\frac{\mathsf{r}^2}{2}\left[ \bar{\cal R}_{\hat 0\hat 1\hat 0\hat 1}\sin^2\theta\cos^2(\phi-\varphi)
 + \bar{\cal R}_{\hat 0\hat 2\hat 0\hat 2} \sin^2\theta\sin^2(\phi-\varphi)+
 \bar{\cal R}_{\hat 0\hat 3\hat 0\hat 3} \cos^2\theta \right]
\ena
where we introduce the spherical coordinates $(\mathsf{r},\theta,\phi)$ in the non-rotating inertial  frame.
This potential can be described by the spherical harmonics 
as 
\bea
\Phi_{\rm tidal}(\mathsf{r},\theta,\phi)
 &=&
 -\mathsf{r}^2 \left\{\frac{1}{2}\sqrt{2\pi\over 15}\left(\bar{\cal R}_{\hat 0\hat 1\hat 0\hat 1}-\bar{\cal R}_{\hat 0\hat 2\hat 0\hat 2}
 \right)\left[Y_{2,2}(\theta,\phi-\varphi)+Y_{2,-2}(\theta,\phi-\varphi)\right]
 \right.
 \nn
 &&
 \left.
 +\frac{1}{3}\sqrt{\pi\over 5}\left(
 2\bar{\cal R}_{\hat 0\hat 3\hat 0\hat 3}-\bar{\cal R}_{\hat 0\hat 1\hat 0\hat 1}-\bar{\cal R}_{\hat 0\hat 2\hat 0\hat 2} \right)Y_{2,0}(\theta)\right\}
 \,,
\ena
where
\beann
Y_{2,0}(\theta)=\sqrt{5\over 16\pi}\left(3\cos^2\theta-1\right)\,,~~
Y_{2,\pm 2}(\theta,\phi)&=&\sqrt{15\over 32\pi}\sin^2\theta e^{\pm 2i\phi}
\,.
\enann

Note that
\bea
&&
\bar{\cal R}_{\hat 0\hat 1\hat 0\hat 1}-\bar{\cal R}_{\hat 0\hat 2\hat 0\hat 2}=-{3\mybar{M}\Delta(\bar{r}_0) \over  F^2_\sigma(\bar{r}_0) \bar{r}_0^3}= -{3\mybar{M} \over  \bar{r}_0^3}\left(1+\delta_1\right)
\\
&&
2\bar{\cal R}_{\hat 0\hat 3\hat 0\hat 3}-\bar{\cal R}_{\hat 0\hat 1\hat 0\hat 1}-\bar{\cal R}_{\hat 0\hat 2\hat 0\hat 2}=
{3\mybar{M} \over  F^2_\sigma(\bar{r}_0) \bar{r}_0^3}\left[\Delta(\bar{r}_0)+(a-\sigma\sqrt{\mybar{M}\bar{r}_0})^2\right]= 
{3\mybar{M} \over  \bar{r}_0^3}\left(1+\delta_1\right)\left(1+\delta_2\right)
\,,
\ena
where
\bea
\delta_1\equiv {S_\sigma(\bar{r}_0) \over  F^2_\sigma(\bar{r}_0) }={\mybar{M}\bar{r}_0+a^2-2a\sigma\sqrt{\mybar{M}\bar{r}_0} \over 
\bar{r}_0^2-3\mybar{M}\bar{r}_0+2a\sigma\sqrt{\mybar{M}\bar{r}_0}}
\,,~~\delta_2\equiv {(a-\sigma\sqrt{\mybar{M}\bar{r}_0})^2\over \bar{r}_0^2-2\mybar{M}\bar{r}_0+a^2}
\label{relativistic_corrections}
\ena
are relativistic corrections by a background Kerr spacetime.

Thus we obtain the tidal potential as
\bea
\Phi_{\rm tidal}
 &=&
\mathsf{r}^2 \left\{\alpha_{20}
Y_{2,0}(\theta,\phi-\varphi)
+\alpha_{22}
\left[Y_{2,2}(\theta,\phi-\varphi)+Y_{2,-2}(\theta,\phi-\varphi)\right]
\right\}
\,,
\label{tidal_potential}
\ena
where
\bea
\alpha_{20}=-\sqrt{\pi\over 5}{\mybar{M} \over  \bar{r}_0^3}\left(1+\delta_1\right)\left(1+\delta_2\right)
\,,~~
\alpha_{22}=\sqrt{3\pi\over 10}{\mybar{M} \over  \bar{r}_0^3}\left(1+\delta_1\right)
\,.
\label{alpha_2022}
\ena

Note that the time dependence of the tidal potential appears through $\varphi(\tau)$.

\subsection{Tidal deformation of a spherical bosonic object}

We first consider a stationary, spherically symmetric solution of the
Schr\"odinger--Poisson system,
\begin{equation}
\psi(\tau, \mathsf{r})=e^{-i\mu \tau}\psi_0(\mathsf{r}),
\end{equation}
where $\mu$ denotes the chemical potential. The equilibrium equations
are
\begin{equation}
\mu\psi_0
=
-\frac{1}{2m_S}\tilde{\nabla}^2\psi_0
+
m_S\Phi_0\psi_0 ,
\end{equation}
and
\begin{equation}
\tilde{\nabla}^2\Phi_0
=
4\pi Gm_S|\psi_0|^2 .
\end{equation}
The spherically symmetric solution of the SP equations can be obtained by numerical integration \cite{MorozPenroseTod1998,SchiveChiuehBroadhurst2014} and approximated by analytic fitting functions \cite{MarshPop2015,TodMoroz1999}.

We then introduce a weak external tidal field generated by the curvature of
the background spacetime. 
We assume that the tidal effect is sufficiently small, i.e., 
\beann
{\Phi_{\rm tidal}\over \Phi_{\rm self}}\sim {{\mybar{M}R_{\rm soliton}^2/\bar{r}_0^3}\over{1/m_S^2 R_{\rm soliton}^2} }
=m_S^2 R_{\rm soliton}^4{   \mybar{M} \over\bar{r}_0^3}=O(\epsilon)\ll  1
\,,
\enann
where $R_{\rm soliton}$ is the size of a soliton,
which must be smaller than the curvature scale  $ \ell_{\bar{\cal R}}\equiv \sqrt{\bar{r}_0^3/\mybar{M}}$ in the present approach.

Since the tidal potential $\Phi_{\rm tidal}$ is described by the spherical harmonics $Y_{2,m} (m=0,\pm 2)$,
we expand the scalar field and gravitational potential in terms of these spherical harmonics 
 as
\begin{equation}
\psi
=
e^{-i\mu \tau}
\left[
\psi_0(\mathsf{r})
+
\psi_2(\tau,\mathsf{r},\theta,\phi)
\right],
\end{equation}
and
\begin{equation}
\Phi_{\rm self}
=
\Phi_0(\mathsf{r})
+
\Phi_2(\tau,\mathsf{r},\theta,\phi)
\,,
\end{equation}
with
\bea
\psi_2&=&\psi_{2, 0}(\tau,\mathsf{r}) Y_{2,0}(\theta,\phi-\varphi)+\psi_{2, 2}(\tau,\mathsf{r}) Y_{2,2}(\theta,\phi-\varphi) 
+ \psi_{2, -2}(\tau,\mathsf{r})Y_{2,-2}(\theta,\phi-\varphi)
\\
\Phi_2&=&\Phi_{2,0}(\tau,\mathsf{r}) Y_{2,0}(\theta,\phi-\varphi)+\Phi_{2,2} (\tau,\mathsf{r}) Y_{2,2}(\theta,\phi-\varphi) 
+ \Phi_{2,-2} (\tau,\mathsf{r})Y_{2,-2}(\theta,\phi-\varphi)
\ena
Here $|\psi_2/\psi_0|$ and $|\Phi_2/\Phi_0|$ are small, of order $O(\epsilon)$, and the higher-order harmonics terms are ignored.

We substitute these expansions into the 
Schr\"odinger equation (\ref{Schrodinger}) as follows:
The left hand side  is 
\beann
&&
i\partial_\tau [e^{-i\mu \tau}(\psi_0+\psi_2)]=e^{-i\mu \tau}\left[\mu(\psi_0+\psi_2)+i\dot \psi_2\right]
\\
&=&
e^{-i\mu \tau}\left[\mu\psi_0+(\mu\psi_{2,0}+i\dot \psi_{2, 0} )Y_{2,0}+
(\mu\psi_{2,2}+i\dot \psi_{2, 2} +2\sigma\mathfrak{w}_{\rm R} \psi_{2, 2})Y_{2,2}+
(\mu\psi_{2,-2}+i\dot \psi_{2, -2} -2\sigma\mathfrak{w}_{\rm R} \psi_{2, -2} )Y_{2,-2}\right]
\,,
\enann
where we have used
\beann
i\dot Y_{2,\pm 2}=\pm 2\sigma  \mathfrak{w}_{\rm R}Y_{2,\pm 2}
\enann

The right hand side of the Schr\"odinger equation is 
\beann
&&
e^{-i\mu \tau}\left[-{\hat \nabla^2\over 2m_S}+m_S(\Phi_{\rm self}+\Phi_{\rm tidal})\right](\psi_0+  \psi_2)
\\
&&
=e^{-i\mu \tau}\left\{
\left(
-{\hat \nabla^2\over 2m_S}+m_S\Phi_0
\right)\psi_0
+ \left[
-{\hat \nabla^2\over 2m_S}\psi_2+m_S\left(
\Phi_0\psi_2+\Phi_2 \psi_0+\Phi_{\rm tidal}\psi_0
\right)\right]
\right\}+O(\epsilon^2)
\enann
Using $\hat{\nabla}^2Y_{2,m}=-{6\over \mathsf{r}^2}Y_{2,m}$,
we find
\beann
&&-{\hat \nabla^2\over 2m_S}\psi_2+m_S\left(
\Phi_0\psi_2+\Phi_2 \psi_0+\Phi_{\rm tidal}\psi_0
\right)
=\left[-{\Delta_{2}\psi_{2,0}\over 2m_S}+m_S \left(\Phi_0\psi_{2,0}
+\psi_0(\Phi_{2,0}+\alpha_{20}\mathsf{r}^2)\right)\right]Y_{2,0}+
\\&&
+\left[-{\Delta_{2}\psi_{2,2}\over 2m_S}+m_S \left(\Phi_0\psi_{2,2}
+\psi_0(\Phi_{2,2}+\alpha_{22}\mathsf{r}^2)\right)\right]Y_{2,2}
+
\left[-{\Delta_{2}\psi_{2,-2}\over 2m_S}+m_S \left(\Phi_0\psi_{2,-2}
+\psi_0(\Phi_{2,-2}+\alpha_{22}\mathsf{r}^2)\right) \right]Y_{2,-2}
\enann
where
$\Delta_{2}\equiv \partial^2_\mathsf{r}+{2\over \mathsf{r}}\partial_\mathsf{r}-{6\over \mathsf{r}^2}$.
We then obtain 
 three perturbation equations as
\bea
i\dot \psi_{2,0} &=&-{\Delta_{2}\psi_{2,0}\over 2m_S}-\mu \psi_{2,0}+m_S\left[\Phi_0 \psi_{2,0}+\left(\Phi_{\rm 2,0}+\alpha_{20}\mathsf{r}^2 \right)\psi_0\right]
\label{Schrodinger_p0}
\\
i\dot \psi_{2,2}&=&-{\Delta_{2}\psi_{2,2}\over 2m_S}-\mu \psi_{2,2}-2\sigma\mathfrak{w}_{\rm R}\psi_{2,2}+m_S\left[\Phi_0 \psi_{2,2}+\left(\Phi_{\rm 2,2}+\alpha_{22}\mathsf{r}^2 \right)\psi_0\right]
\label{Schrodinger_p2}
\\
i\dot \psi_{2,-2}&=&-{\Delta_{2}\psi_{2,-2}\over 2m_S}-\mu \psi_{2,-2}+2\sigma\mathfrak{w}_{\rm R}\psi_{2,-2}+m_S\left[\Phi_0 \psi_{2,-2}+\left(\Phi_{2,-2}+\alpha_{22}\mathsf{r}^2 \right)\psi_0\right]
\label{Schrodinger_p-2}
\ena

The source term of the Poisson equation (\ref{Poisson}) is 
\beann
m_S|\psi|^2= m_S(\psi_0^*\psi_0+\psi_0^*\psi_2+\psi_2^*\psi_0)+O(\epsilon^2)
=m_S\left[|\psi_0|^2+\psi_0(\psi_2+\psi_2^*)\right]+O(\epsilon^2)
\enann
where we choose $\psi_0$ to be real.
Hence three perturbed Poisson equations are 
\bea
\Delta_2 \Phi_{\rm 2,0}&=&4\pi G m_S\psi_0(\psi_{2,0}+\psi_{2,0}^*)
\label{Poisson_p0}
\\
\Delta_2 \Phi_{\rm 2,2}&=&4\pi G m_S\psi_0(\psi_{2,2}+\psi_{2,-2}^*)
\label{Poisson_p2}
\\
\Delta_2 \Phi_{\rm 2,-2}&=&4\pi G m_S\psi_0(\psi_{2,-2}+\psi_{2,2}^*)
\label{Poisson_p-2}
\ena

We have to solve two sets of the coupled differential equations:
\begin{itemize}
  \item the $m=0$ mode [Eq. (\ref{Schrodinger_p0}) and Eq. (\ref{Poisson_p0})], 
  \item the $m=\pm 2$ mode [Eq. (\ref{Schrodinger_p2}), Eq. (\ref{Schrodinger_p-2}), 
  Eq. (\ref{Poisson_p2}) and Eq. (\ref{Poisson_p-2})].
\end{itemize}
\section{Stationary Tidally Deformed Solitons}

We now look for a stationary solution by setting  $\dot \psi_{2,0}=\dot \psi_{2,2}=\dot \psi_{2,-2}=0$.
We obtain the following ordinary differential equations:
\bea
\left({\Delta_{2}\over 2m_S}-m_S\Phi_0+\mu \right)\psi_{2,0}&=&m_S\psi_0\left(\Phi_{\rm 2,0}+\alpha_{20}\mathsf{r}^2 \right)
\label{stationary_p0}
\\
\left({\Delta_{2}\over 2m_S}-m_S\Phi_0 +\mu +2\sigma\mathfrak{w}_{\rm R}\right)\psi_{2,2}&=&m_S\psi_0\left(\Phi_{\rm 2,2}+\alpha_{22}\mathsf{r}^2\right)
\label{stationary_p2}\\
\left({\Delta_{2}\over 2m_S}-m_S\Phi_0 +\mu -2\sigma\mathfrak{w}_{\rm R}\right)\psi_{2,-2}&=&m_S\psi_0\left(\Phi_{\rm 2,-2}+\alpha_{22}\mathsf{r}^2 \right)
\label{stationary_p-2}
\ena
\end{widetext}

Since the coefficients are real, we can choose $\psi_{2,0}\,,\psi_{2,2}$, and $\psi_{2,-2}$ to be real, but $\psi_{2,-2}\neq \psi_{2,2}$
unless $\mathfrak{w}_{\rm R}=0$.
Since the source terms in the Poisson equations (\ref{Poisson_p2}) and (\ref{Poisson_p-2}) are the same, 
we can set 
$ \Phi_{2,-2}=\Phi_{2,2}$. It is because the tidal potential has the same symmetry between $m=2$ and $m=-2$.
The perturbed Poisson equations are 
\bea
\Delta_2 \Phi_{\rm 2,0}&=&8\pi G m_S\psi_0\psi_{2,0}
\label{stationary_Poisson_p0}
\\
\Delta_2 \Phi_{\rm 2,2}&=&4\pi G m_S\psi_0(\psi_{2,2}+\psi_{2,-2})
\label{stationary_Poisson_p2}
\ena
We have two sets of the coupled ordinary differential equations:
\begin{itemize}
  \item the $m=0$ mode [Eq. (\ref{stationary_p0}) and Eq. (\ref{stationary_Poisson_p0})], 
  \item the $m=\pm 2$ mode [Eq. (\ref{stationary_p2}), Eq. (\ref{stationary_p-2}) and 
  Eq. (\ref{stationary_Poisson_p2}) with $ \Phi_{2,-2}=\Phi_{2,2}$].
\end{itemize}
In order to evaluate the observable quantities, we shall introduce dimensionless variables as follows:

Suppose $k$ is a scaling parameter, we set
\bea
\xi&=&k\mathsf{r}
\ena
\bea
\mu&=&{k^2\over m_S}\nu
\\
\psi_0(\mathsf{r})&=&{k^2\over \sqrt{4\pi G m_S^3}}p_0(\xi)
\\
\Phi_0(\mathsf{r})&=&{k^2\over m_S^2}\phi_0(\xi)
\ena
The zeroth-order SP equations for $p_0$ and $\phi_0$ are
\bea
&&
{p_0}''+{2\over \xi}{p_0}'-2(\phi_0- \nu)p_0=0
\\
&&
{\phi_0}''+{2\over \xi}{\phi_0}'=p_0^2
\,,
\ena
where a prime denotes the derivative with respect to $\xi$.
Their boundary conditions are
\bea
&{p_0}'(0)=0\,,~&{p_0}(\infty)=0
\\
&{\phi_0}'(0)=0\,,~&{\phi_0}(\infty)=0
\ena
\\

As for the perturbation variables, 
since the deformation is caused by the tidal force, 
we define the dimensionless variables ${p}_{2,0}, {p}_{2,2} ,{p}_{2,-2}$ and 
$\phi_{\rm 2,0},\phi_{\rm 2,2}$ by
\bea
\psi_{2,0}(\mathsf{r})&=&{k^2\over \sqrt{4\pi G m_S^3}}\epsilon_{\rm tidal}\,{p}_{2,0}(\xi)
\\
\psi_{2,2}(\mathsf{r})&=&{k^2\over \sqrt{4\pi G m_S^3}}\epsilon_{\rm tidal}\,{p}_{2,2}(\xi)
\\
\psi_{2,-2}(\mathsf{r})&=&{k^2\over \sqrt{4\pi G m_S^3}}\epsilon_{\rm tidal}\,{p}_{2,-2}(\xi)
\ena
\bea
\Phi_{2,0}(r)&=&{k^2\over m_S^2}\epsilon_{\rm tidal} \phi_{\rm 2,0}(\xi)
\\
\Phi_{2,2}(r)&=&{k^2\over m_S^2}\epsilon_{\rm tidal} \phi_{\rm 2,2}(\xi)
\,,
\ena
where
\bea
\epsilon_{\rm tidal}\equiv {m_S^2\over k^4}{\mybar M\over \bar{r}_0^3}
\ena
is the perturbation parameter due to the tidal force.

We find two sets of the perturbation equations as:\\
\begin{widetext}
$\langle$The $m=0$ mode : $p_{2,0}$ and $\phi_{2,0}$ $\rangle$
\bea
{\varDelta}_{2}p_{2,0}+2\left(\nu -\phi_0\right)p_{2,0}&=&
2\left[\phi_{2,0}-\sqrt{\pi\over 5}\xi^2 (1+\delta_1)(1+\delta_2)\right]p_0
\label{stationary_tp0}
\\
 \varDelta_2 \phi_{\rm 2,0}&=&2\phi_0 p_{2,0 }
\label{Poisson_tp0}
\ena
$\langle$The $m=\pm 2$ mode : $p_{2,2}, p_{2,-2}$ and $\phi_{2,2}$ $\rangle$
\bea
{\varDelta}_{2}p_{2,2}+2\left(\nu -\phi_0+2\sigma\varpi_{\rm R}\right)p_{2,2}&=&
2\left[\phi_{2,2}+\sqrt{3\pi\over 10}\xi^2 (1+\delta_1)\right]p_0
\label{stationary_tp2}
\\
{\varDelta}_{2}p_{2,-2}+2\left(\nu -\phi_0-2\sigma\varpi_{\rm R}\right)p_{2,-2}&=&
2\left[\phi_{2,2}+\sqrt{3\pi\over 10}\xi^2 (1+\delta_1)\right]p_0
\label{stationary_tp-2}
\\
 \varDelta_2 \phi_{2,2}&=& p_0 (p_{2,2}+p_{2,-2})
\label{Poisson_tp2}
\,,
\ena
\end{widetext}
where
\bea
{\varDelta}_{2}&\equiv&{d^2\over d\xi^2}+2{d\over d\xi}-{6\over \xi^2}
\,,
\\
\varpi_{\rm R}&\equiv &{m_S \over k^2}\mathfrak{w}_{\rm R}
\,.
\ena

In what follows, without numerically solving the above equations, we estimate the relevant physical quantities and discuss their potential observability.

\section{Tidal deformability, Love number and gravitational waves}

\subsection{Quadrupole moment and Love number}
As shown in the previous section, the soliton deformation is time-dependent in the non-rotating inertial frame, as described by the spherical harmonics ($Y_{2,\pm 2}(\theta,\phi-\varphi)$). However, in the rotating inertial frame with coordinates ($\hat{\vect{x}}=(\mathsf{r},\theta,\hat{\phi}\equiv\phi-\varphi)$), the deformation is static. The deformed soliton is thus stationary in the rotating frame and rotates with angular velocity $\mathfrak{w}_{\rm R}$ in the non-rotating inertial frame.

The quadrupolar deformation induced by an external tidal field can be
characterized by the induced mass quadrupole moment $Q_{\hat i\hat j}$
 in the rotating inertial  frame,
which
 is defined by
\begin{equation}
Q_{\hat i\hat j}
=
\int d^3\hat{\vect{x}}\,
\rho_S(\hat{\vect{x}})
\left(
x_{\hat i} x_{\hat j}
-\frac{1}{3}\delta_{\hat i\hat j}\mathsf{r}^2
\right).
\end{equation}

Using the perturbative expansion,
\beann
\rho_S
&= &
m_S(\psi_0+\psi_2)^*(\psi_0+\psi_2)
\\
&= &
m_S\left[|\psi_0|^2+\psi_0(\psi_2^*+\psi_2)\right]+O(\epsilon^2)
\\
&=&\rho_{S,0}(\mathsf{r})+m_S\psi_0\Big{[}2\psi_{2,0} Y_{2,0}
\\
&&+(\psi_{2,2}+\psi_{2,-2})(Y_{2,2}+Y_{2,-2})\Big{]}+O(\epsilon^2)
\,,
\enann
after the angular direction is integrated, we obtain the quadrupole moment in the rotating inertial frame as
\beann
Q_{xx}-Q_{yy}&=&4\sqrt{2\pi\over 15}m_S\int d\mathsf{r} \mathsf{r}^4\psi_0(\psi_{2,2}+\psi_{2,-2})
\\
Q_{zz}&=&{8\over 3}\sqrt{\pi\over 5}m_S\int d\mathsf{r}  \mathsf{r}^4 \psi_0\psi_{2,0}
\enann

 In the linear response regime, the relation between the quadrupole moment 
 $Q_{\hat i\hat j}$ and
the external tidal field ${\cal E}_{\hat i\hat j}$ is written as
\bea
Q_{\hat i\hat j}=-\lambda_{\rm Love}{\cal E}_{\hat i\hat j}
\ena
where $\lambda_{\rm Love}$ is the tidal deformability, which is related to the Love number.
${\cal E}_{\hat i\hat  j}$ is given by the tidal potential
(\ref{tidal_potential}).

We then obtain
\beann
\lambda_{\rm Love}^{(0)}&\equiv&-{Q_{zz}\over {\cal E}_{zz}}=-{8\over 3\alpha_{20}}\sqrt{\pi\over 5}m_S\int d\mathsf{r} \mathsf{r}^4 \psi_0\psi_{2,0}
\\
&=&{8m_S\bar{r}_0^3\over 3\mybar{M}(1+\delta_1)(1+\delta_2)}\int d\mathsf{r} \mathsf{r}^4 \psi_0\psi_{2,0}
\nn
&=&{2\over 3\pi G k^5(1+\delta_1)(1+\delta_2)}\int d\xi \xi^4 p_0 p_{2,0}
\\
\lambda_{\rm Love}^{(2)}&\equiv&-{Q_{xx}-Q_{yy}\over {\cal E}_{xx}-{\cal E}_{yy}}
\\
&=&-{4\over \alpha_{22}}
\sqrt{2\pi\over 15}m_S\int d\mathsf{r} \mathsf{r}^4\psi_0(\psi_{2,2}+\psi_{2,-2})
\\
&=&-{8m_S\bar{r}_0^3\over 3\mybar{M}(1+\delta_1)}\int d\mathsf{r} \mathsf{r}^4\psi_0(\psi_{2,2}+\psi_{2,-2})
\nn
&=&-{2\over 3\pi Gk^5(1+\delta_1)}\int d\xi \xi^4 p_0 (p_{2,2}+p_{2,-2})
\enann

We define the dimensionless quadrupolar Love number as
\begin{equation}
\mathsf{k}_2
=
\frac{3G}{2}
\frac{\lambda}{R^5},
\end{equation}
where $R$ represents the characteristic radius of the bosonic object.
Hence, up to relativistic corrections and numerical factors arising from the integration of the dimensionless solution, the present model gives
\beann
\mathsf{k}_2\sim \frac{1}{\pi (R_{\rm soliton}k)^5}=\frac{1}{\pi \Xi_{\rm soliton}^5}
\enann
where $\Xi_{\rm soliton}$ is the radius of the dimensionless soliton solution.
The scaling estimate suggests that $\mathsf{k}_2$ may be of order unity, although its numerical value must be determined by solving the perturbation equations.

The present Schr\"odinger--Poisson framework provides a direct way to
calculate these quantities for weakly relativistic bosonic objects.
In contrast to neutron stars, where the equation of state determines
the tidal response, the response of bosonic objects is controlled by the
wave nature of the scalar field and the balance between gravitational
attraction and quantum pressure.

\subsection{Gravitational Waves}
The soliton rotates about its own center of mass with an angular frequency $\mathfrak{w}_{\rm R}$ in the non-rotating inertial frame. The initially spherical soliton is deformed by the tidal force exerted by the black hole, acquiring a quadrupole moment ($Q_{\hat i\hat j}$). This deformation is expected to lead to the emission of gravitational waves. We estimate the gravitational-wave emission using the following simplified approach:

The non-axisymmetric quadrupole is associated with the $m=\pm2$ harmonics, and 
it rotates with the angular frequency $2\mathfrak{w}_{\rm R}$. 
Using $\ddot Q\sim 4\mathfrak{w}_{\rm R}^2 Q$, 
the amplitude of the gravitational waves for the observer at the distance $D$ is estimated as
\bea
h_{\rm GW}\sim {2G\over D}\ddot Q\sim {8G\over D}\mathfrak{w}_{\rm R}^2Q\sim 
{8G\over D}\mathfrak{w}_{\rm R}^2\lambda_{\rm Love}^{(2)}{\cal E}
\,.
\ena
Since ${\cal E}$ is evaluated by the tidal force of the black hole at the radius $\bar r_0$ as
$
{\cal E}\sim {\mybar M/\bar{r}_0^3}
$,
we  obtain 
\beann
h_{\rm GW}\sim  
{8G\mathfrak{w}_{\rm R}^2\lambda_{\rm Love}^{(2)}\mybar M\over D\bar{r}_0^3}
\enann

Using $\mathfrak{w}_{\rm R}=\sqrt{\mybar M/\bar{r}_0^3}$ and 
$\lambda_{\rm Love}^{(2)}={2\over 3G}\mathsf{k}_2^{(2)}R_{\rm soliton}^5$, where $\mathsf{k}_2^{(2)}$ is the dimensionless Love number associated with the $m=\pm2$ response, 
we find
\bea
h_{\rm GW}&\sim &
{16\over 3}\mathsf{k}_2^{(2)}{\mybar M\over D}\left({\Mbar\over \bar{r}_0}\right)^6
\left({R_{\rm soliton}\over \Mbar}\right)^5
\label{GW_h}
\ena
To assess the observability, we adopt the reference 
 observational constraint ($h_{\rm GW}\geq 10^{-22}$) by LISA,  
 which
  gives
\bea
{R_{\rm soliton}\over \Mbar}\geq 2.85\times 10^{-5} \left({D\over \Mbar}\right)^{1\over 5}\left({\bar{r}_0\over \Mbar}\right)^{6\over 5}{\mathsf{k}_2^{(2)}}^{-{1\over 5}}
\ena

Suppose that a soliton is orbiting around the center of our Galaxy, i.e., $M=4\times 10^6 M_\odot$, $D=8.2 {\rm kpc}$,
which gives $\mybar M/D\approx 2.33\times 10^{-11}$.
Assuming $\mathsf{k}_2^{(2)}\sim O(1)$, we find 
\beann
{R_{\rm soliton}\over \Mbar}\geq 3.8\times 10^{-3} \left({\bar{r}_0\over \Mbar}\right)^{6\over 5}
\enann
or
\bea
{R_{\rm soliton}\over \bar{r}_0}\geq 3.8\times 10^{-3} 
\left({\bar{r}_0\over \Mbar}\right)^{1\over 5}
\,.
\label{size_constraint}
\ena

The frequency of gravitational waves is $f_{\rm GW}=\mathfrak{w}_{\rm R}/\pi$.
Assuming the relevant frequency range for LISA is $10^{-4}{\rm Hz}\leq f\leq 1 {\rm Hz}$, we find the constraint for $\bar{r}_0$ is
\bea
r_{\rm ISCO}(a)\leq \bar{r}_0\leq 30\Mbar
\label{frequency_constraint}
\,,
\ena
where $r_{\rm ISCO}(a)$ is the ISCO radius of the Kerr 
rotation parameter $a$. 
The lower bound $r_{\rm ISCO}(a)$ arises from the 
orbital stability condition.

\begin{figure}[htbp]
\begin{center}
\includegraphics[width=4.2cm]{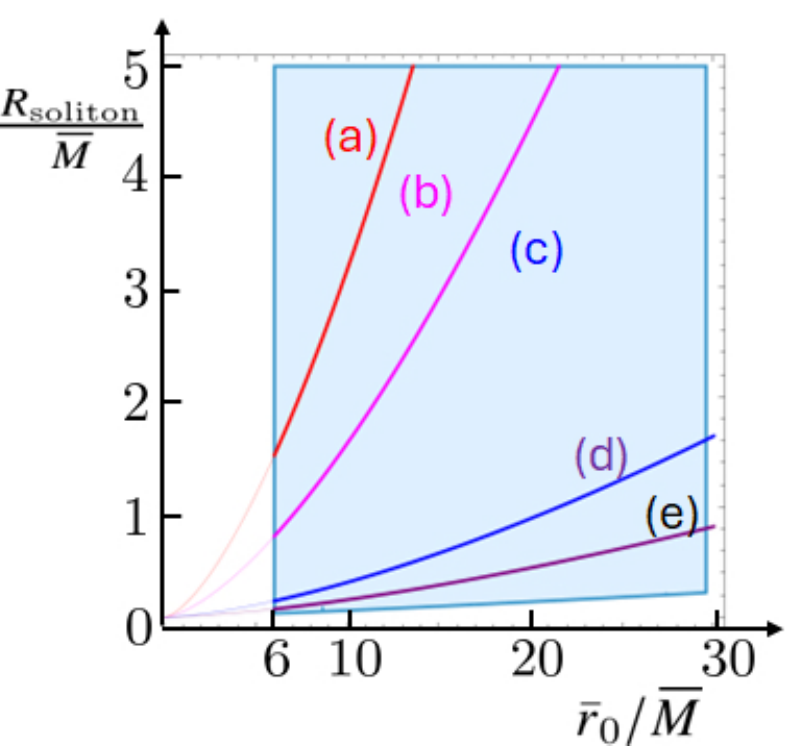}
\includegraphics[width=4.2cm]{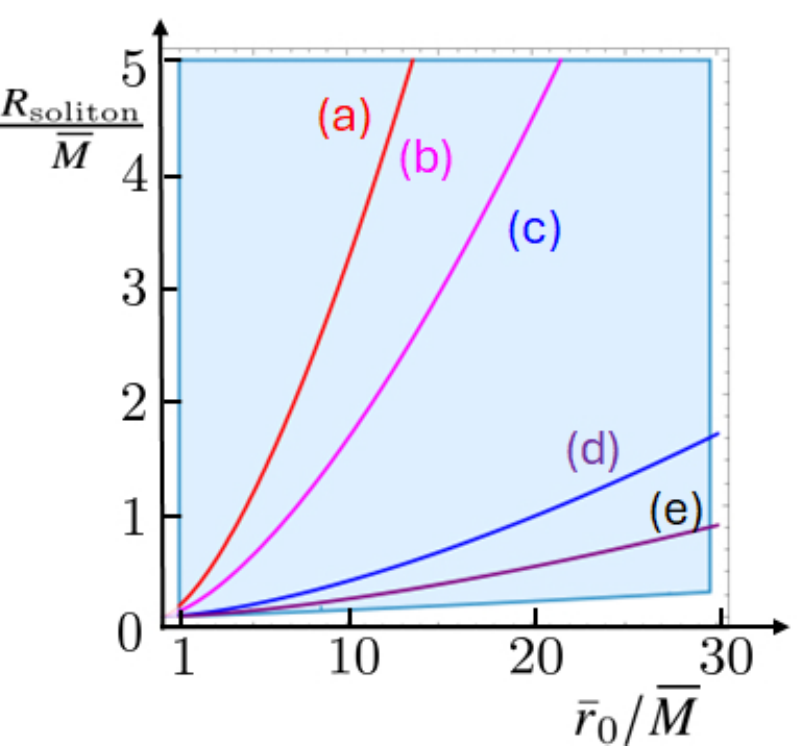}
\\
~~~~(i) \hskip 4cm (ii)
\caption{The light-blue region indicates the parameter space in which both constraints, Eqs. (\ref{size_constraint}) and (\ref{frequency_constraint}), are satisfied. The left panel (i) and the right panel (ii) correspond to the Schwarzschild ($a=0$) and extreme Kerr ($a= M$) black holes, respectively.
The red (a), magenta (b), blue (c), and purple (d)  
curves correspond to 
$R_{\rm soliton}/\ell_{\cal \bar{R}}=0.1$, $0.05$, $0.01$, and $0.005$, respectively.}
\label{Soliton_GW}
\end{center}
\end{figure}

In Fig. \ref{Soliton_GW}, we show the light-blue region where both constraints (\ref{size_constraint}) and (\ref{frequency_constraint}) are satisfied.
The left panel (i) and right panel (ii) correspond to the Schwarzschild 
and extreme Kerr black holes, respectively. 
We also show four models with $R_{\rm soliton}/\ell_{\bar{\cal R}}= 0.1, 0.05, 0.01$ and $0.005$ by 
(a) red, (b) magenta, (c) blue,  and (d) purple curves, respectively. 
Note that our model is valid for $R_{\rm soliton}\ll \ell_{\bar{\cal R}}$.

\begin{figure}[htbp]
\begin{center}
\includegraphics[width=8cm]{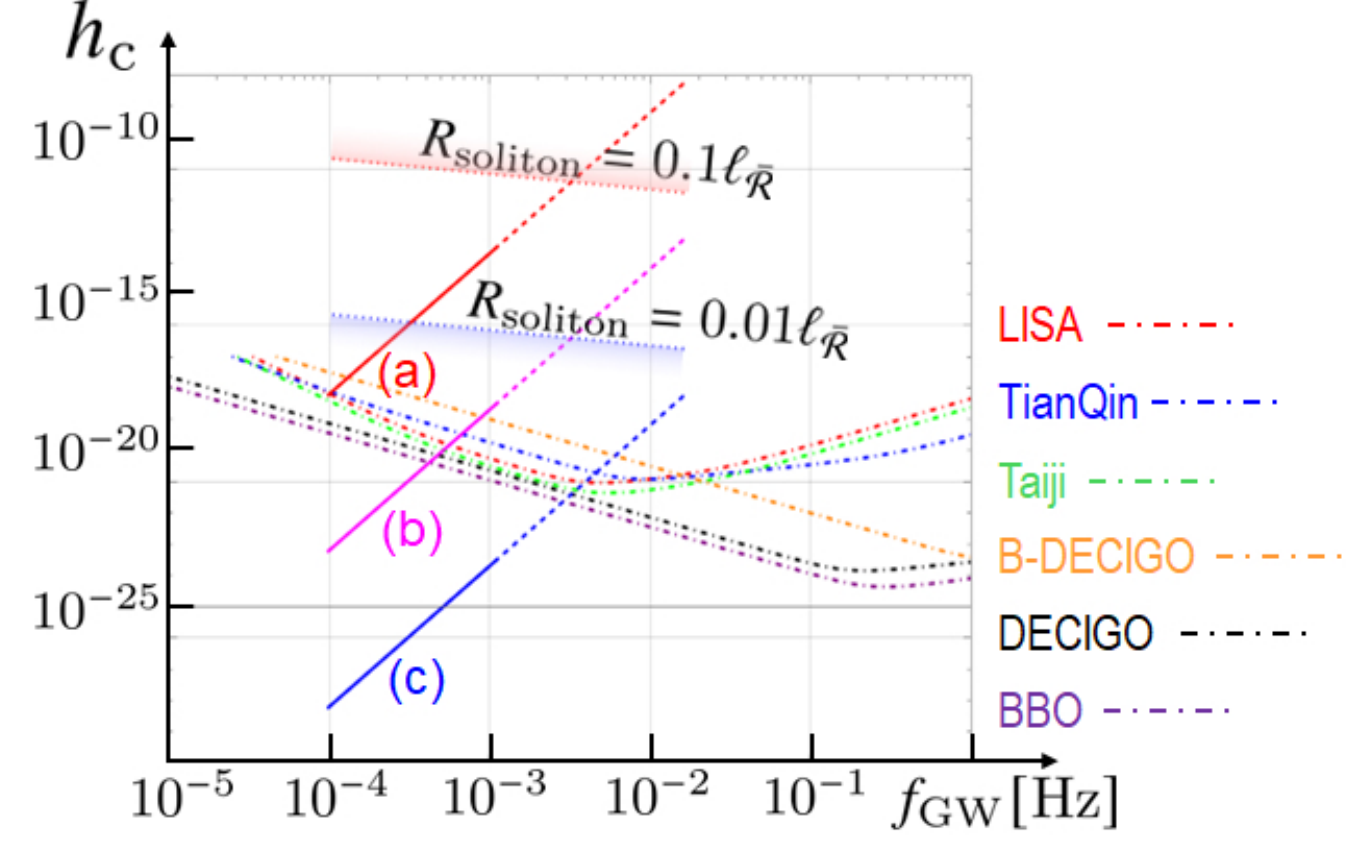}
\caption{The estimated characteristic strain of gravitational waves associated with tidally deformed
solitons for the range of Eq. (\ref{frequency_constraint}) with sizes (a) $R_{\rm soliton}=GM$ (red lines)\,,
(b) $R_{\rm soliton}=0.1 GM$ (magenta lines)\,, and
(c) $R_{\rm soliton}=0.01GM$ (blue lines).
The solid lines represent the results for Schwarzschild black hole ($a=0$) and overlap with the corresponding parts of those for the extreme Kerr black hole ($a=M$), which are shown by both solid and dashed lines.
The detector sensitivity curves for LISA, TianQin, Taiji, B-DECIGO,
DECIGO, and BBO are also shown
\cite{Robson2019,Luo2016,Ruan2018,Isoyama2018,Yagi2011,Yagi2017Erratum,Kawamura2021}.
The cases $R_{\rm soliton}=0.1\ell_{\cal \bar{R}}$ and
$R_{\rm soliton}=0.01\ell_{\cal \bar{R}}$ are included for reference,
although the present approach is formally valid for
$R_{\rm soliton}\ll \ell_{\cal \bar{R}}$.}
\label{chracteristic_strain_GW_soliton}
\end{center}
\end{figure}

In order to compare the detector sensitivity, we shall discuss the characteristic strain $h_c$,
which is naively evaluated as
\beann
h_c=h_{\rm GW}\sqrt{f_{\rm GW}T_{\rm obs}}
\enann 
where $T_{\rm obs}$ is the observation period.
Assuming that $T_{\rm obs}\approx  4 {\rm years}$, in Fig. \ref{chracteristic_strain_GW_soliton},
 we show the characteristic strain of the gravitational waves 
emitted from a deformed soliton and the detector sensitivity curves 
 for LISA, TianQin, Taiji, B-DECIGO, DECIGO, and BBO \cite{Robson2019,Luo2016,Ruan2018,Isoyama2018,Yagi2011,Yagi2017Erratum,Kawamura2021}.
The sizes of deformed solitons are chosen as
 (a) $R_{\rm soliton}=GM$  (red lines)\,, $R_{\rm soliton}=0.1 GM$ (magenta lines)\,, and  (c) $R_{\rm soliton}=0.01 GM$ (blue lines).
The solid lines represent the results for Schwarzschild black hole ($a=0$) and overlap with the corresponding parts of those for the extreme Kerr black hole ($a=M$), which are shown by both solid and dashed lines.
In the present 
simplified treatment, only the upper frequency bound, which is coming from the ISCO radius, depends on the spin parameter $a$.

We also show two curves, $R_{\rm soliton}=0.1\ell_{\cal \bar{R}}$
 (the red dot-dashed curve) and $R_{\rm soliton}=0.01\ell_{\cal \bar{R}}$ (the blue dot-dashed curve), for reference.
Since the present approach is valid for $R_{\rm soliton}\ll\ell_{\cal \bar{R}}$, the models in the region below the blue dot-dashed curve are appropriate, whereas those in the region above the red dot-dashed curve may require extending the metric expansion in a local inertial frame 
to include higher-order terms involving derivatives of the Riemann curvature tensor
\cite{IshiiShibataMino2005}. 
The validity of the approximation is less certain in the intermediate region.
Our estimates suggest that these gravitational waves could be detectable by future space-based gravitational-wave detectors.
 
\section{CONCLUSION AND REMARKS}

In this work, we have formulated a curved-spacetime extension of the
Schr\"odinger-Poisson (SP) description of a self-gravitating bosonic
soliton. Our formulation is based on a local inertial frame centered
on the soliton and provides a systematic way of incorporating the
curvature of the background spacetime into the non-relativistic
dynamics.

We have shown that, in the weak-field and non-relativistic regime, the
effect of the external gravitational field appears as an effective
tidal potential in the Schr\"odinger equation. By imposing the
center-of-mass condition, the local frame can be chosen to be
non-rotating inertial frame and the center of mass of the soliton follows 
a geodesic at the order considered here. The
resulting SP equations describe the internal dynamics of the soliton
in the presence of the tidal field of the background spacetime.

We have then considered a soliton moving on a circular equatorial
geodesic around a Kerr black hole. Using a local inertial frame
constructed along the orbit, we have derived the relevant components of the
Riemann curvature tensor and obtained the corresponding tidal
potential. In the non-rotating inertial frame, the tidal field contains
a stationary component as well as time-dependent components
due to a circular motion of a soliton. This provides a natural
framework for describing the deformation of a self-gravitating
bosonic soliton in the vicinity of a black hole.

We have also discussed stationary rotating solitons
 in the presence of a tidal field.
We have estimated the induced quadrupole moment,
 and  obtained the corresponding expressions for the quadrupole
response and 
 the dimensionless Love number $\mathsf{k}_2$. 
These quantities provide
a characterization of the response of the self-gravitating bosonic
configuration to the external tidal field.

As a possible observational consequence, we have considered
gravitational radiation associated with the rotating soliton. We
have obtained the characteristic scaling as Eq. (\ref{GW_h}).
It shows the strong dependence of the signal on the size of the
soliton $R_{\rm soliton}$ and on its orbital radius $\bar r_0$.
 We have also estimated the
corresponding characteristic strain and compared it with
representative sensitivity curves of future gravitational-wave
detectors. These estimates identify parameter regions in which the predicted characteristic strain may fall within the sensitivity ranges of future space-based gravitational-wave detectors. 

In these analyses, we have neglected relativistic effects associated with the background spacetime, 
such as those represented by 
$\delta_1$ and $\delta_2$, as well as redshift corrections 
to the gravitational waves due to the motion of the soliton in the deep gravitational field.
 This is because we have not numerically solved the dimensionless equations, and these effects may 
 also change our present results by numerical factors.

Several extensions remain to be investigated. First, a detailed
numerical study of the stationary SP solutions obtained here is
needed to determine the Love number and tidal deformability for
specific soliton configurations. 
It is also important to examine
the stability of these stationary solutions by introducing small
perturbations around them and following their subsequent time
evolution. Such an analysis will clarify whether the stationary
tidally deformed configurations are dynamically stable and will
provide information on the characteristic internal modes of the
soliton.
A detailed numerical investigation of the stationary solutions,
their stability, tidal Love numbers, fully dynamical response, and
the resulting gravitational-wave waveforms will be presented
elsewhere.

A further extension is to consider more general 
orbits, including eccentric and unbound scattering orbits. Such
systems provide a natural setting for studying not only the
dynamical tidal response of the soliton but also the conditions under
which the soliton remains gravitationally bound or is tidally
disrupted (see \cite{Edwards_2018}). The resulting dynamics may also leave characteristic
signatures in the emitted gravitational waves.

Beyond the present perturbative approach, 
a fully time-dependent treatment of
the SP system is an important direction for future work. In such a
treatment, the time dependence of the tidal field and the internal
dynamics of the soliton can be followed simultaneously. This may
reveal dynamical tidal effects, excitation of internal modes, and
resonant phenomena that cannot be captured by the stationary
solutions. Such effects may also modify the gravitational-wave
signal and its frequency structure.

The framework developed here is not restricted to a particular
realization of a bosonic compact object. Whenever an appropriate
non-relativistic description is applicable, it can be used to study
different self-gravitating scalar configurations, including boson
stars, axion stars, and other solitonic objects. The present work
thus provides a framework connecting the internal structure of
self-gravitating bosonic configurations, their tidal response in
curved spacetime, and their possible gravitational-wave signatures.

\begin{acknowledgments}
K.M. thanks Richard Easther for valuable discussions.
K.M. would also like to thank the Department of Physics at the University of Auckland 
for its hospitality and support, which made it possible to carry out part of this work during his stay.
This work was supported in part by the JSPS KAKENHI
Grant Number JP24K07058. 
\end{acknowledgments}
 
 \newpage
~
 \newpage
\appendix

\section{Proper reference frame of an arbitrary observer
in a curved spacetime}
\label{proper reference frame}

In this appendix, we discuss how to set up a local inertial frame in a curved spacetime
\cite{Misner1973,Manasse1963,Gorbatsievich_Bobrik}. 
Consider a freely falling observer in an external gravitational field.
Its orbit  is given by 
a world line $ \gamma$ described by 
\beann
x^\mu=z^\mu(\tau)
\,,
\enann
where $\tau$ is a proper time of the observer. 
The 4-velocity is given by
\beann
u^\mu(\tau)\equiv {dz^\mu\over d\tau}
\,.
\enann

We then set up an orthonormal tetrad system 
$\{e_{\hat \alpha}^\mu\}$ along the observer's world line $\gamma$, which is defined  by
the conditions such that
\beann
e_{\hat \alpha}^{~\mu} e_{\hat \beta \mu}=\eta_{\hat \alpha \hat \beta}\,,~~e_{\hat 0}^{~\mu}=u^\mu\,,
\enann
where $\eta_{\hat \alpha \hat \beta}$ is 
the Minkowski spacetime metric.

For a given 4-velocity $u^\mu$,
 this tetrad system is determined up to three-dimensional rotations.
The tetrad $e_{\hat j}^\mu$  is transported  along the observer's world line  $ \gamma$ 
as
\beann
{D e_{\hat j}^{~\mu}\over d\tau}=-\Omega^{\mu\nu}\,e_{\hat j \,\nu}
\,,
\enann
where 
\beann
\Omega^{\mu\nu}\equiv a^\mu  u^\nu-u^\mu a^\nu  +u_\alpha \omega_\beta \epsilon^{\alpha\beta\mu\nu}
\,.
\enann
Here,
\beann
a^\mu&\equiv& {Du^\mu\over d\tau},
\enann
and 
\beann
\omega_\mu&\equiv&{1\over 2}
u^\alpha\epsilon_{\alpha\mu\rho\sigma}\Omega^{\rho\sigma}={1\over 2}\stackrel{(3)}{\epsilon}_{\mu\rho\sigma}\Omega^{\rho\sigma},
\enann
are  the acceleration of the observer
and  the angular velocity of a rotating 
spatial basis vector $e_{\hat j}^\mu$, respectively. A non-rotating tetrad frame for which $\omega^\mu=0$ is said to be Fermi--Walker transported.
If the orbit is a geodesic ($a^\mu=0$ and $\omega^\mu=0$), we find 
$\displaystyle{{D e_{\hat j}^\mu/ d\tau}=0}$, which is just a parallel transport.

Next, we construct a local coordinate system (the observer's proper reference system) 
near the world line $\gamma$, which is described as
\beann
(x^{\hat\mu})=(\tau, x^{\hat a})\,,
\enann
where the spatial components $x^{\hat a}$ are measured from the point at $\tau$ on the world line $\gamma$ along the spatial hypersurface $\Sigma(\tau)$ perpendicular to $\gamma$.

We introduce the acceleration and angular velocity in the proper reference frame, which are  defined by
$
a^{\hat j} \equiv   e^{\hat j}_{~\mu} a^\mu \,,~
\omega^{\hat j} \equiv  e^{\hat j}_{~\mu}\omega^\mu
\,,
$ respectively.
We also denote the tetrad component of the Riemann curvature and the angular velocity tensor
by $\bar{\cal R}_{\hat \mu\hat\nu\hat\rho\hat \sigma}$
and
$
\omega_{\hat j\hat k}\equiv \epsilon_{\hat j \hat k \hat \ell}\omega^{\hat \ell}
$, respectively.
Using these variables, we find the metric form of this proper reference frame 
up to the second order of $x^{\hat a}$ 
as~\cite{Misner1973,Manasse1963,Gorbatsievich_Bobrik}.
\bea
g_{\hat \mu\hat \nu}=\eta_{\hat \mu \hat \nu}+\varepsilon_{\hat \mu \hat \nu}
+O(|x^{\hat k}|^3),
\label{metric_reference_frame}
\ena 
\begin{widetext}
where
\bea
\varepsilon_{\hat 0 \hat 0}&=&-\left[2a_{\hat k}
x^{\hat k}
+\left(
\bar{\cal R}_{\hat 0 \hat k \hat 0 \hat \ell }-
\omega_{\hat j\hat k} \omega^{\hat j}_{~\hat \ell} \right)
x^{\hat k}  x^{\hat \ell} + \left(a_{\hat k} x^{\hat k}\right)^2   \right],
\label{epsilon00}\\
\varepsilon_{\hat 0 \hat j}&=&- \left[ \omega_{\hat j\hat k} x^{\hat k} 
 +{2\over 3}  \bar{\cal R}_{\hat 0\hat k \hat j  \hat \ell }x^{\hat k}x^{\hat \ell}\right],
\label{epsilon0j}\\
\varepsilon_{\hat i \hat j}&=&- {1\over 3} \bar{\cal R}_{\hat i \hat k \hat j\hat \ell}x^{\hat k}x^{\hat \ell}
\,.
\label{epsilonij}
\ena

\end{widetext}

For the Fermi-Walker transport, $\omega^{\hat k}=0$, 
which we call a non-rotating inertial frame. 
In addition, if $a^{\hat j}=0$, the observer follows a geodesic,
i.e., the tetrad is parallel-transported along the world line.
We recover the Fermi normal coordinates.

\newpage
~
\newpage

\section{Riemann curvature components in the local inertial frame}
\label{Riemann_curvature}

\subsection{Circular motion in Kerr spacetime}

We consider a circular geodesic on the equatorial plane 
of a test particle with a unit mass in Kerr spacetime (\ref{Kerr_metric}).
The proper energy and proper angular momentum at the radius $\bar r_0$
are determined by the radius $ \bar{r}_0$ as
\bea
E&=&{\bar{r}_0^2-2\mybar M\bar{r}_0+a\sigma \sqrt{\mybar M\bar{r}_0}\over \bar{r}_0 F_\sigma(\bar{r}_0)},
\label{energy}
\\
L&=& {\sigma\sqrt{\mybar M\bar{r}_0}
(\bar{r}_0^2+a^2- 2a\sigma\sqrt{\mybar M\bar{r}_0})\over \bar{r}_0 F_\sigma(\bar{r}_0)}
\,,
\label{angular_momentum}
\ena
respectively, 
where 
\beann
F_\sigma(\bar{r}_0)\equiv \left(\bar{r}_0^2-3\mybar M\bar{r}_0+ 2a\sigma\sqrt{\mybar M\bar{r}_0}\right)^{1/2},
\enann
and
$\sigma=+1$ or $-1$, which 
 correspond to prograde and retrograde orbits, respectively.
The existence condition of a circular orbit is $F_\sigma(\bar{r}_0)>0$.

Since the energy and angular momentum are defined by
\beann
E=-u_{\mathfrak{t}}
\,,~
L=u_{\phi}
\,,
\enann
we find
\beann
u^{\mathfrak{t}}&=&{\bar{r}_0^2+a\sigma\sqrt{\mybar M\bar{r}_0}\over \bar{r}_0 F_\sigma(\bar{r}_0)},
\\
u^{\phi}&=&{\sigma\sqrt{\mybar M\bar{r}_0}\over 
\bar{r}_0 F_\sigma(\bar{r}_0)}.
\enann
From the latter equation, 
we find  the angular velocity of a circular observer measured by the proper time as
\beann
 {d\phi\over d\tau}=u^{\phi}=\sigma  \mathfrak{w}_0,
 \enann
 where
 \beann
 \mathfrak{w}_0\equiv
{\sqrt{\mybar M\bar{r}_0}\over 
\bar{r}_0 F_\sigma(\bar{r}_0)}
\,.
\enann

\subsection{A local inertial reference frame}
For the present purpose, there is one convenient  tetrad system of  Kerr spacetime,
which is called Carter's tetrad system such that
\beann
e_{~\bar{\mathfrak{t}}}^\mu&=&{1\over \sqrt{\Sigma\Delta}}\left(r^2+a^2, 0, 0, a\right),
\\
e_{~\bar{\mathfrak{r}}}^\mu&=&\sqrt{\Delta\over \Sigma}\left(0,1,0,0\right),
\\
e_{~\bar \theta}^\mu&=& {1\over\sqrt{\Sigma}}\left(0, 0, 1, 0\right),
\\
e_{~\bar \phi}^\mu&=&{1\over \sqrt{\Sigma}}\left(a\sin\theta, 0, 0, {1\over \sin\theta}\right)
\,.
\enann

Now we construct a local inertial reference frame along the observer.
The transformation matrix from Carter's tetrad to 
a rotating proper reference frame $(\hat x^\mu)=(\tau, x, y, z)$
is given by
\beann
\Lambda_{\hat 0}^{~\bar \alpha}&=&\left({E(\bar{r}_0^2+a^2)-aL\over \bar{r}_0\sqrt{\Delta}}\,,0\,,0\,,{L-aE\over \bar{r}_0}\right),\\
\Lambda_{\hat x}^{~\bar \alpha}&=&{1\over \sqrt{S}}\left(0\,,
\bar{r}_0\Lambda_{\hat 0}^{~\bar 0}\,,0\,,0\right),
\\
\Lambda_{\hat y}^{~\bar \alpha}&=&\sigma\sqrt{K\over S}\left(\Lambda_{\hat 0}^{~\bar 0}\,,0\,,0\,,{S\over K}\Lambda_{\hat 0}^{~\bar \phi}\right),
\\
\Lambda_{\hat z}^{~\bar \alpha}&=&
{\sigma\over \sqrt{K}}\left(0\,,0\,,-\bar{r}_0
\Lambda_{\hat 0}^{~\bar \phi}\,,0\right)
\,,
\enann
where
\beann
K&\equiv&\left(L-aE\right)^2,
\\
S&\equiv&\bar{r}_0^2+K
\,.
\enann
$K$ is the so-called Carter's constant.

\begin{figure}[htbp]
\begin{center}
\includegraphics[width=4cm]{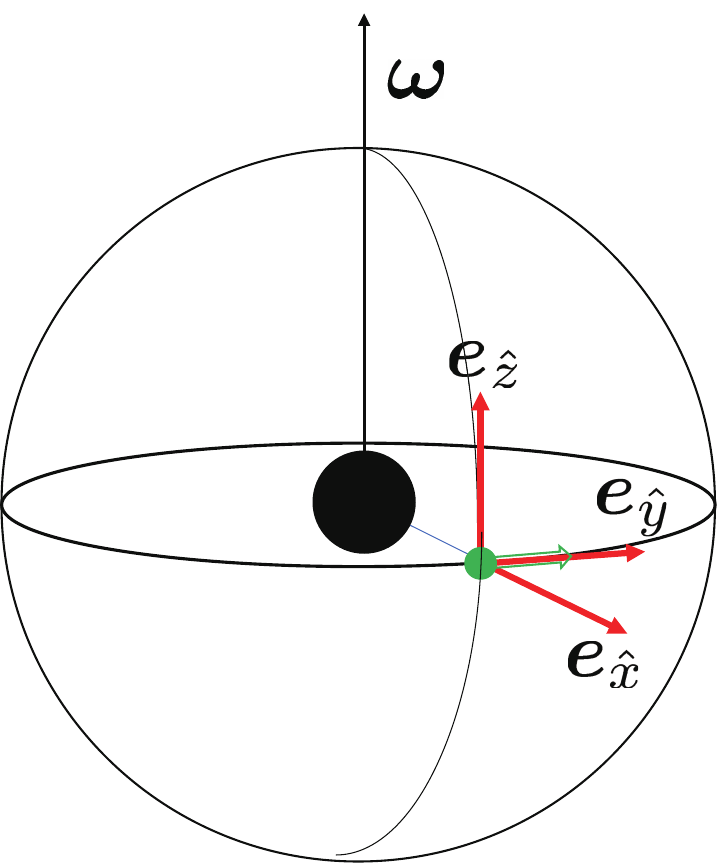}
\caption{A tetrad system $\{
\vect{e}_{\hat x}\,, \vect{e}_{\hat y}\,, \vect{e}_{\hat z} \}$ rotating with an angular velocity $\mathfrak{w}_0$ along a circular orbit. }
\label{tetrad}
\end{center}
\end{figure}

In the rotating proper reference frame, the basis vectors in the $x,y,z$ directions are chosen as Fig. \ref{tetrad}.

Inserting Eqs. (\ref{energy}) and (\ref{angular_momentum}),
we obtain the transformation matrix $\Lambda_{\hat \alpha}^{~\bar \lambda}$ as
\bea
\Lambda_{\hat 0}^{~\bar \lambda}&=&
{1\over F_\sigma(\bar{r}_0)}\left(\sqrt{\Delta(\bar{r}_0)}
\,, 0\,, 0\,,\sigma\sqrt{\mybar{M}\bar{r}_0}-a\right),
~~~~~
\label{Lambda_hat0}\\
\Lambda_{\hat x}^{~\bar \lambda}&=& 
\left(0\,,1\,, 0\,,0\right),
\label{Lambda_hatx}\\
\Lambda_{\hat y}^{~\bar \lambda}&=& {1\over F_\sigma(\bar{r}_0)}
\left(\sigma\sqrt{\mybar{M}\bar{r}_0}-a
\,, 0\,, 0\,,\sqrt{\Delta(\bar{r}_0)}
\right),~~~~~~~~
\label{Lambda_haty}\\
\Lambda_{\hat z}^{~\bar \lambda}&=& \left(0\,, 0\,, -1\,,0\right)
\,,
\label{Lambda_hatz}
\ena
which provides the tetrad of the rotating proper reference frame $e_{\hat \alpha}^{~\mu}\equiv 
\Lambda_{\hat \alpha}^{~\bar \lambda}e_{\bar \lambda}^{~\mu}
$ as
\begin{widetext}
\beann
e_{\hat 0}^{~\mu}&\equiv&
{1\over \bar{r}_0F_\sigma(\bar{r}_0)}\left(\bar{r}_0^2+a\sigma\sqrt{\mybar M\bar{r}_0}
\,, 0\,, 0\,,\sigma\sqrt{\mybar M\bar{r}_0}\right),
\\
e_{\hat x}^{~\mu}&=& 
\left(0\,,{\sqrt{\Delta(\bar{r}_0)}\over \bar{r}_0}\,, 0\,,0\right),
\\
e_{\hat y}^{~\mu}&=& {1\over \bar{r}_0 F_\sigma(\bar{r}_0)\sqrt{\Delta(\bar{r}_0)}}
\left(\sigma\sqrt{\mybar M\bar{r}_0}\left[\bar{r}_0^2+a^2-2a\sigma\sqrt{\mybar M\bar{r}_0}\right]\,, 0\,, 0\,,
\bar{r}_0^2-2\mybar M\bar{r}_0+a\sigma\sqrt{\mybar M\bar{r}_0}\right),
\\
e_{\hat z}^{~\mu}&=& \left(0\,, 0\,, -{1\over \bar{r}_0}\,,0\right)
\,.
\enann
 \end{widetext}

Since we are interested in a non-rotating proper reference frame with $\omega^\mu=0$,
we have to rotate the $xy$-plane as
($(x,y) \rightarrow (\mathsf{x},\mathsf{y})$)
to find a non-rotating local inertial frame with the coordinate 
$(\tilde x^\mu)=(\tau, \mathsf{x},\mathsf{y},\mathsf{z})$.
The  transformation is given by 
\beann
\tilde \Lambda_{\hat {\mathsf{x}}}^{~\bar \alpha}
&=&\Lambda_{\hat x}^{~\bar \alpha}\, \cos \varphi (\tau)-\Lambda_{\hat y}^{~\bar \alpha}\,\sin\varphi(\tau),
\\
\tilde \Lambda_{\hat{\mathsf{y}}}^{~\bar \alpha}
&=&\Lambda_{\hat x}^{~\bar \alpha}\, \sin \varphi (\tau)+\Lambda_{\hat y}^{~\bar \alpha}\,\cos\varphi(\tau)
\,,
\enann
where the rotation angle $\varphi$ satisfies 
the evolution equation such that
\beann
\dot{\varphi}&=&
\sigma
{\sqrt{K}\over \bar{r}_0^2}\left({E(\bar{r}_0^2+a^2)-aL\over S}
+{a(L-aE)\over K}
\right)
\\
&
\equiv& \sigma\mathfrak{w}_{\rm R}
\,,
\enann
with
\beann
\mathfrak{w}_{\rm R} ={\mybar M^{1/2}\over \bar{r}_0^{3/2}}\,.
\enann
The rotation angle $\varphi$ is now 
\beann
\varphi=\sigma \mathfrak{w}_R\,\tau
\,.
\enann

\subsection{Riemann curvature components in local inertial frame}
The non-trivial components of the Riemann curvature 
in the Carter tetrad system  on the equatorial plane are given by
\beann
&&
R_{\bar{t}\bar \theta\bar{t}\bar \theta}
\,=\,R_{\bar{t}\bar \phi\bar{t}\bar \phi}
\,=\,-R_{\bar{r}\bar \theta\bar{r}\bar \theta}
\,=\,-R_{\bar{r}\bar \phi\bar{r}\bar \phi}
\,=\, {\mybar M\over \bar r^3}
\,,
\\
&&
R_{\bar \theta\bar \phi\bar \theta\bar \phi}
\,=\,
-R_{\bar{t}\bar{r}\bar{t}\bar{r}}
\,=\,{2\mybar M\over \bar r^3}
\,.
\label{Riemann_Carter}
\enann

Using the transformation matrices  $\Lambda_{\hat \alpha}^{~\bar \lambda}$ 
and $\tilde \Lambda_{\hat \alpha}^{~\tilde \beta}$, we find the  Riemann curvature 
components in the rotating inertial frame and non-rotating  inertial frame.
The non-trivial  Riemann curvature components in the rotating inertial frame are given by
\beann
\bar{\cal R}_{\hat 0 \hat 1 \hat 0 \hat 1}
&=&
-\bar{\cal R}_{\hat 2 \hat 3 \hat 2 \hat 3}
=-{\mybar M\over  \bar{r}_0^3}{G^2_\sigma(\bar{r}_0)\over  F^2_\sigma(\bar{r}_0) }
\,,~
\\
\bar{\cal R}_{\hat 0 \hat 2 \hat 0 \hat 2}&=&-\bar{\cal R}_{\hat 3\hat 1\hat 3 \hat 1}={\mybar M\over \bar{r}_0^3}
\,,~
\\
\bar{\cal R}_{\hat 0\hat 3\hat 0\hat 3}&=&-\bar{\cal R}_{\hat 1\hat 2 \hat 1\hat 2}
={\mybar M\over  \bar{r}_0^3}{H^2_\sigma(\bar{r}_0)\over  F^2_\sigma(\bar{r}_0) },
\\
\bar{\cal R}_{\hat 0 \hat1\hat 2\hat 1}&=&-\bar{\cal R}_{\hat 0 \hat 3\hat 2\hat 3}
=-{3 \mybar M\over  \bar{r}_0^3}{\left(\sigma\sqrt{\mybar M\bar{r}_0}-a\right)\sqrt{\Delta(\bar{r}_0)}
\over  F^2_\sigma(\bar{r}_0) }
\,,
\enann
where
\beann
G^2_\sigma(\bar{r}_0)&\equiv&2\bar{r}_0^2+3a^2-3\mybar M\bar{r}_0-2a\sigma\sqrt{\mybar M\bar{r}_0}
\\
H^2_\sigma(\bar{r}_0)&\equiv&\bar{r}_0^2+3a^2-4a\sigma\sqrt{\mybar M\bar{r}_0}
\enann
and those 
 in the non-rotating inertial frame are 
 \beann
 \bar{\cal R}_{\tilde 0 \tilde 1 \tilde 0 \tilde 1} &=&\bar{\cal R}_{\hat 0\hat 1\hat 0\hat 1}\cos^2\varphi +\bar{\cal R}_{\hat 0\hat 2\hat 0\hat 2}\sin^2\varphi
 \\
\bar{\cal R}_{\tilde 0 \tilde 2 \tilde 0 \tilde 2} &=&\bar{\cal R}_{\hat 0\hat 1\hat 0\hat 1}\sin^2\varphi+\bar{\cal R}_{\hat 0\hat 2\hat 0\hat 2}\cos^2\varphi
 \\
\bar{\cal R}_{\tilde 0 \tilde 3 \tilde 0 \tilde 3} &=&\bar{\cal R}_{\hat 0\hat 3\hat 0\hat 3}
 \\
\bar{\cal R}_{\tilde 0 \tilde 1 \tilde 0 \tilde 2} &=&
\left(
\bar{\cal R}_{\hat 0\hat 1\hat 0\hat 1}-\bar{\cal R}_{\hat 0\hat 2\hat 0\hat 2}\right)\cos\varphi\sin\varphi
\\
\bar{\cal R}_{\tilde 0 \tilde 2 \tilde 0 \tilde 3} &=&\bar{\cal R}_{\hat 0\hat 1\hat 0\hat 3}
\sin\varphi
\nonumber \\
\bar{\cal R}_{\tilde 0 \tilde 3 \tilde 0 \tilde 1} &=&\bar{\cal R}_{\hat 0\hat 1\hat 0\hat 3}
\cos\varphi
\nonumber \,.
 \enann
 
\newpage
\bibliography{soliton}

\begin{thebibliography}{40}
\expandafter\ifx\csname natexlab\endcsname\relax\def\natexlab#1{#1}\fi
\expandafter\ifx\csname bibnamefont\endcsname\relax
  \def\bibnamefont#1{#1}\fi
\expandafter\ifx\csname bibfnamefont\endcsname\relax
  \def\bibfnamefont#1{#1}\fi
\expandafter\ifx\csname citenamefont\endcsname\relax
  \def\citenamefont#1{#1}\fi
\expandafter\ifx\csname url\endcsname\relax
  \def\url#1{\texttt{#1}}\fi
\expandafter\ifx\csname urlprefix\endcsname\relax\def\urlprefix{URL }\fi
\providecommand{\bibinfo}[2]{#2}
\providecommand{\eprint}[2][]{\url{#2}}

\bibitem[{\citenamefont{Bertone et~al.}(2005)\citenamefont{Bertone, Hooper, and Silk}}]{BertoneHooperSilk2005}
\bibinfo{author}{\bibfnamefont{G.}~\bibnamefont{Bertone}}, \bibinfo{author}{\bibfnamefont{D.}~\bibnamefont{Hooper}}, \bibnamefont{and} \bibinfo{author}{\bibfnamefont{J.}~\bibnamefont{Silk}}, \bibinfo{journal}{Phys. Rep.} \textbf{\bibinfo{volume}{405}}, \bibinfo{pages}{279} (\bibinfo{year}{2005}).

\bibitem[{\citenamefont{Carr and K{\"u}hnel}(2020)}]{CarrKuhnel2020}
\bibinfo{author}{\bibfnamefont{B.}~\bibnamefont{Carr}} \bibnamefont{and} \bibinfo{author}{\bibfnamefont{F.}~\bibnamefont{K{\"u}hnel}}, \bibinfo{journal}{Ann. Rev. Nucl. Part. Sci.} \textbf{\bibinfo{volume}{70}}, \bibinfo{pages}{355} (\bibinfo{year}{2020}).

\bibitem[{\citenamefont{Kaup}(1968)}]{Kaup1968}
\bibinfo{author}{\bibfnamefont{D.~J.} \bibnamefont{Kaup}}, \bibinfo{journal}{Phys. Rev.} \textbf{\bibinfo{volume}{172}}, \bibinfo{pages}{1331} (\bibinfo{year}{1968}).

\bibitem[{\citenamefont{Ruffini and Bonazzola}(1969)}]{RuffiniBonazzola1969}
\bibinfo{author}{\bibfnamefont{R.}~\bibnamefont{Ruffini}} \bibnamefont{and} \bibinfo{author}{\bibfnamefont{S.}~\bibnamefont{Bonazzola}}, \bibinfo{journal}{Phys. Rev.} \textbf{\bibinfo{volume}{187}}, \bibinfo{pages}{1767} (\bibinfo{year}{1969}).

\bibitem[{\citenamefont{Liebling and Palenzuela}(2023)}]{LieblingPalenzuela2023}
\bibinfo{author}{\bibfnamefont{S.~L.} \bibnamefont{Liebling}} \bibnamefont{and} \bibinfo{author}{\bibfnamefont{C.}~\bibnamefont{Palenzuela}}, \bibinfo{journal}{Living Reviews in Relativity} \textbf{\bibinfo{volume}{26}}, \bibinfo{pages}{1} (\bibinfo{year}{2023}).

\bibitem[{\citenamefont{Braaten and Zhang}(2019)}]{BraatenZhang2019}
\bibinfo{author}{\bibfnamefont{E.}~\bibnamefont{Braaten}} \bibnamefont{and} \bibinfo{author}{\bibfnamefont{H.}~\bibnamefont{Zhang}}, \bibinfo{journal}{Rev. Mod. Phys.} \textbf{\bibinfo{volume}{91}}, \bibinfo{pages}{041002} (\bibinfo{year}{2019}).

\bibitem[{\citenamefont{Page}(2004)}]{Page2004}
\bibinfo{author}{\bibfnamefont{D.~N.} \bibnamefont{Page}}, \bibinfo{journal}{Phys. Rev. D} \textbf{\bibinfo{volume}{70}}, \bibinfo{pages}{023002} (\bibinfo{year}{2004}).

\bibitem[{\citenamefont{Hui et~al.}(2017)\citenamefont{Hui, Ostriker, Tremaine, and Witten}}]{HuiOstrikerTremaineWitten2017}
\bibinfo{author}{\bibfnamefont{L.}~\bibnamefont{Hui}}, \bibinfo{author}{\bibfnamefont{J.~P.} \bibnamefont{Ostriker}}, \bibinfo{author}{\bibfnamefont{S.}~\bibnamefont{Tremaine}}, \bibnamefont{and} \bibinfo{author}{\bibfnamefont{E.}~\bibnamefont{Witten}}, \bibinfo{journal}{Phys. Rev. D} \textbf{\bibinfo{volume}{95}}, \bibinfo{pages}{043541} (\bibinfo{year}{2017}).

\bibitem[{\citenamefont{Hui}(2021)}]{Hui2021}
\bibinfo{author}{\bibfnamefont{L.}~\bibnamefont{Hui}}, \bibinfo{journal}{Annual Review of Astronomy and Astrophysics} \textbf{\bibinfo{volume}{59}}, \bibinfo{pages}{247} (\bibinfo{year}{2021}).

\bibitem[{\citenamefont{Braaten et~al.}(2016)\citenamefont{Braaten, Mohapatra, and Zhang}}]{BraatenMohapatraZhang2016NR}
\bibinfo{author}{\bibfnamefont{E.}~\bibnamefont{Braaten}}, \bibinfo{author}{\bibfnamefont{A.}~\bibnamefont{Mohapatra}}, \bibnamefont{and} \bibinfo{author}{\bibfnamefont{H.}~\bibnamefont{Zhang}}, \bibinfo{journal}{Phys. Rev. D} \textbf{\bibinfo{volume}{94}}, \bibinfo{pages}{076004} (\bibinfo{year}{2016}).

\bibitem[{\citenamefont{Aoki et~al.}(2018)\citenamefont{Aoki, Maeda, Misonoh, and Okawa}}]{Aoki:2017ixz}
\bibinfo{author}{\bibfnamefont{K.}~\bibnamefont{Aoki}}, \bibinfo{author}{\bibfnamefont{K.}~\bibnamefont{Maeda}}, \bibinfo{author}{\bibfnamefont{Y.}~\bibnamefont{Misonoh}}, \bibnamefont{and} \bibinfo{author}{\bibfnamefont{H.}~\bibnamefont{Okawa}}, \bibinfo{journal}{Phys. Rev. D} \textbf{\bibinfo{volume}{97}}, \bibinfo{pages}{044005} (\bibinfo{year}{2018}), \eprint{1710.05606}.

\bibitem[{\citenamefont{Bucciotti and Trincherini}(2023)}]{BucciottiTrincherini2023}
\bibinfo{author}{\bibfnamefont{B.}~\bibnamefont{Bucciotti}} \bibnamefont{and} \bibinfo{author}{\bibfnamefont{E.}~\bibnamefont{Trincherini}}, \bibinfo{journal}{J. High Energy Phys.} \textbf{\bibinfo{volume}{2023}}, \bibinfo{pages}{193} (\bibinfo{year}{2023}), \eprint{2309.02482}.

\bibitem[{\citenamefont{Kadota et~al.}(2024)\citenamefont{Kadota, Kim, Ko, and Yang}}]{KadotaKimKoYang2024}
\bibinfo{author}{\bibfnamefont{K.}~\bibnamefont{Kadota}}, \bibinfo{author}{\bibfnamefont{J.~H.} \bibnamefont{Kim}}, \bibinfo{author}{\bibfnamefont{P.}~\bibnamefont{Ko}}, \bibnamefont{and} \bibinfo{author}{\bibfnamefont{X.-Y.} \bibnamefont{Yang}}, \bibinfo{journal}{Phys. Rev. D} \textbf{\bibinfo{volume}{109}}, \bibinfo{pages}{015022} (\bibinfo{year}{2024}).

\bibitem[{\citenamefont{Hinderer}(2008)}]{Hinderer2008}
\bibinfo{author}{\bibfnamefont{T.}~\bibnamefont{Hinderer}}, \bibinfo{journal}{Astrophys. J.} \textbf{\bibinfo{volume}{677}}, \bibinfo{pages}{1216} (\bibinfo{year}{2008}), \eprint{0711.2420}.

\bibitem[{\citenamefont{Binnington and Poisson}(2009)}]{BinningtonPoisson2009}
\bibinfo{author}{\bibfnamefont{T.}~\bibnamefont{Binnington}} \bibnamefont{and} \bibinfo{author}{\bibfnamefont{E.}~\bibnamefont{Poisson}}, \bibinfo{journal}{Phys. Rev. D} \textbf{\bibinfo{volume}{80}}, \bibinfo{pages}{084018} (\bibinfo{year}{2009}), \eprint{0906.1366}.

\bibitem[{\citenamefont{Damour and Nagar}(2009)}]{Damour_2009}
\bibinfo{author}{\bibfnamefont{T.}~\bibnamefont{Damour}} \bibnamefont{and} \bibinfo{author}{\bibfnamefont{A.}~\bibnamefont{Nagar}}, \bibinfo{journal}{Phys. Rev. D} \textbf{\bibinfo{volume}{80}} (\bibinfo{year}{2009}), ISSN \bibinfo{issn}{1550-2368}.

\bibitem[{\citenamefont{Flanagan and Hinderer}(2008)}]{FlanaganHinderer2008}
\bibinfo{author}{\bibfnamefont{{\'E}.~{\'E}.} \bibnamefont{Flanagan}} \bibnamefont{and} \bibinfo{author}{\bibfnamefont{T.}~\bibnamefont{Hinderer}}, \bibinfo{journal}{Phys. Rev. D} \textbf{\bibinfo{volume}{77}}, \bibinfo{pages}{021502} (\bibinfo{year}{2008}), \eprint{0709.1915}.

\bibitem[{\citenamefont{Cardoso et~al.}(2017)\citenamefont{Cardoso, Franzin, Maselli, Pani, and Raposo}}]{CardosoFranzinMaselliPaniRaposo2017}
\bibinfo{author}{\bibfnamefont{V.}~\bibnamefont{Cardoso}}, \bibinfo{author}{\bibfnamefont{E.}~\bibnamefont{Franzin}}, \bibinfo{author}{\bibfnamefont{A.}~\bibnamefont{Maselli}}, \bibinfo{author}{\bibfnamefont{P.}~\bibnamefont{Pani}}, \bibnamefont{and} \bibinfo{author}{\bibfnamefont{G.}~\bibnamefont{Raposo}}, \bibinfo{journal}{Phys. Rev. D} \textbf{\bibinfo{volume}{95}}, \bibinfo{pages}{084014} (\bibinfo{year}{2017}).

\bibitem[{\citenamefont{Sennett et~al.}(2017)\citenamefont{Sennett, Hinderer, Steinhoff, Buonanno, and Ossokine}}]{SennettHindererSteinhoffBuonannoOssokine2017}
\bibinfo{author}{\bibfnamefont{N.}~\bibnamefont{Sennett}}, \bibinfo{author}{\bibfnamefont{T.}~\bibnamefont{Hinderer}}, \bibinfo{author}{\bibfnamefont{J.}~\bibnamefont{Steinhoff}}, \bibinfo{author}{\bibfnamefont{A.}~\bibnamefont{Buonanno}}, \bibnamefont{and} \bibinfo{author}{\bibfnamefont{S.}~\bibnamefont{Ossokine}}, \bibinfo{journal}{Phys. Rev. D} \textbf{\bibinfo{volume}{96}}, \bibinfo{pages}{024002} (\bibinfo{year}{2017}).

\bibitem[{\citenamefont{Chen et~al.}(2024)\citenamefont{Chen, Huang, Zhao, and Wang}}]{ChenHuangZhaoWang2024}
\bibinfo{author}{\bibfnamefont{J.-R.} \bibnamefont{Chen}}, \bibinfo{author}{\bibfnamefont{L.-X.} \bibnamefont{Huang}}, \bibinfo{author}{\bibfnamefont{L.}~\bibnamefont{Zhao}}, \bibnamefont{and} \bibinfo{author}{\bibfnamefont{Y.-Q.} \bibnamefont{Wang}}, \bibinfo{journal}{Phys. Rev. D} \textbf{\bibinfo{volume}{109}}, \bibinfo{pages}{104078} (\bibinfo{year}{2024}).

\bibitem[{\citenamefont{{Manasse} and {Misner}}(1963)}]{Manasse1963}
\bibinfo{author}{\bibfnamefont{F.~K.} \bibnamefont{{Manasse}}} \bibnamefont{and} \bibinfo{author}{\bibfnamefont{C.~W.} \bibnamefont{{Misner}}}, \bibinfo{journal}{J. Math. Phys.} \textbf{\bibinfo{volume}{4}}, \bibinfo{pages}{735} (\bibinfo{year}{1963}).

\bibitem[{\citenamefont{Gorbatsievich and Bobrik}(2010)}]{Gorbatsievich_Bobrik}
\bibinfo{author}{\bibfnamefont{A.}~\bibnamefont{Gorbatsievich}} \bibnamefont{and} \bibinfo{author}{\bibfnamefont{A.}~\bibnamefont{Bobrik}}, \bibinfo{journal}{AIP Conference Proceedings} \textbf{\bibinfo{volume}{1205}}, \bibinfo{pages}{87} (\bibinfo{year}{2010}).

\bibitem[{\citenamefont{Ishii et~al.}(2005)\citenamefont{Ishii, Shibata, and Mino}}]{IshiiShibataMino2005}
\bibinfo{author}{\bibfnamefont{M.}~\bibnamefont{Ishii}}, \bibinfo{author}{\bibfnamefont{M.}~\bibnamefont{Shibata}}, \bibnamefont{and} \bibinfo{author}{\bibfnamefont{Y.}~\bibnamefont{Mino}}, \bibinfo{journal}{Phys. Rev. D} \textbf{\bibinfo{volume}{71}}, \bibinfo{pages}{044017} (\bibinfo{year}{2005}).

\bibitem[{\citenamefont{Maeda et~al.}(2023{\natexlab{a}})\citenamefont{Maeda, Gupta, and Okawa}}]{Maeda:2023tao}
\bibinfo{author}{\bibfnamefont{K.}~\bibnamefont{Maeda}}, \bibinfo{author}{\bibfnamefont{P.}~\bibnamefont{Gupta}}, \bibnamefont{and} \bibinfo{author}{\bibfnamefont{H.}~\bibnamefont{Okawa}}, \bibinfo{journal}{Phys. Rev. D} \textbf{\bibinfo{volume}{107}}, \bibinfo{pages}{124039} (\bibinfo{year}{2023}{\natexlab{a}}).

\bibitem[{\citenamefont{Maeda et~al.}(2023{\natexlab{b}})\citenamefont{Maeda, Gupta, and Okawa}}]{Maeda:2023uyx}
\bibinfo{author}{\bibfnamefont{K.}~\bibnamefont{Maeda}}, \bibinfo{author}{\bibfnamefont{P.}~\bibnamefont{Gupta}}, \bibnamefont{and} \bibinfo{author}{\bibfnamefont{H.}~\bibnamefont{Okawa}}, \bibinfo{journal}{Phys. Rev. D} \textbf{\bibinfo{volume}{108}}, \bibinfo{pages}{123041} (\bibinfo{year}{2023}{\natexlab{b}}).

\bibitem[{\citenamefont{Maeda and Okawa}(2025)}]{Maeda:2025row}
\bibinfo{author}{\bibfnamefont{K.}~\bibnamefont{Maeda}} \bibnamefont{and} \bibinfo{author}{\bibfnamefont{H.}~\bibnamefont{Okawa}}, \bibinfo{journal}{Phys. Rev. D} \textbf{\bibinfo{volume}{112}}, \bibinfo{pages}{044014} (\bibinfo{year}{2025}), \eprint{2504.18934}.

\bibitem[{\citenamefont{Maeda and Okawa}(2026)}]{Maeda:2026}
\bibinfo{author}{\bibfnamefont{K.}~\bibnamefont{Maeda}} \bibnamefont{and} \bibinfo{author}{\bibfnamefont{H.}~\bibnamefont{Okawa}} (\bibinfo{year}{2026}), \eprint{2605.22525}.

\bibitem[{\citenamefont{Moroz et~al.}(1998)\citenamefont{Moroz, Penrose, and Tod}}]{MorozPenroseTod1998}
\bibinfo{author}{\bibfnamefont{I.~M.} \bibnamefont{Moroz}}, \bibinfo{author}{\bibfnamefont{R.}~\bibnamefont{Penrose}}, \bibnamefont{and} \bibinfo{author}{\bibfnamefont{P.}~\bibnamefont{Tod}}, \bibinfo{journal}{Classical and Quantum Gravity} \textbf{\bibinfo{volume}{15}}, \bibinfo{pages}{2733} (\bibinfo{year}{1998}).

\bibitem[{\citenamefont{Schive et~al.}(2014)\citenamefont{Schive, Chiueh, and Broadhurst}}]{SchiveChiuehBroadhurst2014}
\bibinfo{author}{\bibfnamefont{H.-Y.} \bibnamefont{Schive}}, \bibinfo{author}{\bibfnamefont{T.}~\bibnamefont{Chiueh}}, \bibnamefont{and} \bibinfo{author}{\bibfnamefont{T.}~\bibnamefont{Broadhurst}}, \bibinfo{journal}{Nature Physics} \textbf{\bibinfo{volume}{10}}, \bibinfo{pages}{496} (\bibinfo{year}{2014}).

\bibitem[{\citenamefont{Marsh and Pop}(2015)}]{MarshPop2015}
\bibinfo{author}{\bibfnamefont{D.~J.~E.} \bibnamefont{Marsh}} \bibnamefont{and} \bibinfo{author}{\bibfnamefont{A.-R.} \bibnamefont{Pop}}, \bibinfo{journal}{Monthly Notices of the Royal Astronomical Society} \textbf{\bibinfo{volume}{451}}, \bibinfo{pages}{2479} (\bibinfo{year}{2015}), \eprint{1502.03456}.

\bibitem[{\citenamefont{Tod and Moroz}(1999)}]{TodMoroz1999}
\bibinfo{author}{\bibfnamefont{P.}~\bibnamefont{Tod}} \bibnamefont{and} \bibinfo{author}{\bibfnamefont{I.~M.} \bibnamefont{Moroz}}, \bibinfo{journal}{Nonlinearity} \textbf{\bibinfo{volume}{12}}, \bibinfo{pages}{201} (\bibinfo{year}{1999}).

\bibitem[{\citenamefont{Robson et~al.}(2019)\citenamefont{Robson, Cornish, and Liu}}]{Robson2019}
\bibinfo{author}{\bibfnamefont{T.}~\bibnamefont{Robson}}, \bibinfo{author}{\bibfnamefont{N.}~\bibnamefont{Cornish}}, \bibnamefont{and} \bibinfo{author}{\bibfnamefont{C.}~\bibnamefont{Liu}}, \bibinfo{journal}{Class. Quantum Grav.} \textbf{\bibinfo{volume}{36}}, \bibinfo{pages}{105011} (\bibinfo{year}{2019}), \eprint{1803.01944}.

\bibitem[{\citenamefont{Luo et~al.}(2016)\citenamefont{Luo, Chen, Duan, Gong, Hu, Ji, Liu, Mei, Milyukov, Sazhin et~al.}}]{Luo2016}
\bibinfo{author}{\bibfnamefont{J.}~\bibnamefont{Luo}}, \bibinfo{author}{\bibfnamefont{L.-S.} \bibnamefont{Chen}}, \bibinfo{author}{\bibfnamefont{H.-Z.} \bibnamefont{Duan}}, \bibinfo{author}{\bibfnamefont{Y.-G.} \bibnamefont{Gong}}, \bibinfo{author}{\bibfnamefont{S.-C.} \bibnamefont{Hu}}, \bibinfo{author}{\bibfnamefont{J.-H.} \bibnamefont{Ji}}, \bibinfo{author}{\bibfnamefont{Q.}~\bibnamefont{Liu}}, \bibinfo{author}{\bibfnamefont{J.}~\bibnamefont{Mei}}, \bibinfo{author}{\bibfnamefont{V.}~\bibnamefont{Milyukov}}, \bibinfo{author}{\bibfnamefont{M.}~\bibnamefont{Sazhin}}, \bibnamefont{et~al.}, \bibinfo{journal}{Class. Quantum Grav.} \textbf{\bibinfo{volume}{33}}, \bibinfo{pages}{035010} (\bibinfo{year}{2016}).

\bibitem[{\citenamefont{Ruan et~al.}(2018)\citenamefont{Ruan, Guo, Cai, and Zhang}}]{Ruan2018}
\bibinfo{author}{\bibfnamefont{W.-H.} \bibnamefont{Ruan}}, \bibinfo{author}{\bibfnamefont{Z.-K.} \bibnamefont{Guo}}, \bibinfo{author}{\bibfnamefont{R.-G.} \bibnamefont{Cai}}, \bibnamefont{and} \bibinfo{author}{\bibfnamefont{Y.-Z.} \bibnamefont{Zhang}}, \bibinfo{journal}{International Journal of Modern Physics A} \textbf{\bibinfo{volume}{33}}, \bibinfo{pages}{1843016} (\bibinfo{year}{2018}), \eprint{1807.09495}.

\bibitem[{\citenamefont{Isoyama et~al.}(2018)\citenamefont{Isoyama, Nakano, and Nakamura}}]{Isoyama2018}
\bibinfo{author}{\bibfnamefont{S.}~\bibnamefont{Isoyama}}, \bibinfo{author}{\bibfnamefont{H.}~\bibnamefont{Nakano}}, \bibnamefont{and} \bibinfo{author}{\bibfnamefont{T.}~\bibnamefont{Nakamura}}, \bibinfo{journal}{Prog. Theor. Exp. Phys.} \textbf{\bibinfo{volume}{2018}}, \bibinfo{pages}{073E01} (\bibinfo{year}{2018}), \eprint{1802.06977}.

\bibitem[{\citenamefont{Yagi and Seto}(2011)}]{Yagi2011}
\bibinfo{author}{\bibfnamefont{K.}~\bibnamefont{Yagi}} \bibnamefont{and} \bibinfo{author}{\bibfnamefont{N.}~\bibnamefont{Seto}}, \bibinfo{journal}{Phys. Rev. D} \textbf{\bibinfo{volume}{83}}, \bibinfo{pages}{044011} (\bibinfo{year}{2011}), \eprint{1101.3940}.

\bibitem[{\citenamefont{Yagi and Seto}(2017)}]{Yagi2017Erratum}
\bibinfo{author}{\bibfnamefont{K.}~\bibnamefont{Yagi}} \bibnamefont{and} \bibinfo{author}{\bibfnamefont{N.}~\bibnamefont{Seto}}, \bibinfo{journal}{Phys. Rev. D} \textbf{\bibinfo{volume}{95}}, \bibinfo{pages}{109901} (\bibinfo{year}{2017}).

\bibitem[{\citenamefont{Kawamura et~al.}(2021)\citenamefont{Kawamura, Ando, Seto, Sato et~al.}}]{Kawamura2021}
\bibinfo{author}{\bibfnamefont{S.}~\bibnamefont{Kawamura}}, \bibinfo{author}{\bibfnamefont{M.}~\bibnamefont{Ando}}, \bibinfo{author}{\bibfnamefont{N.}~\bibnamefont{Seto}}, \bibinfo{author}{\bibfnamefont{S.}~\bibnamefont{Sato}}, \bibnamefont{et~al.}, \bibinfo{journal}{Prog. Theor. Exp. Phys.} \textbf{\bibinfo{volume}{2021}}, \bibinfo{pages}{05A105} (\bibinfo{year}{2021}), \eprint{2006.13545}.

\bibitem[{\citenamefont{Edwards et~al.}(2018)\citenamefont{Edwards, Kendall, Hotchkiss, and Easther}}]{Edwards_2018}
\bibinfo{author}{\bibfnamefont{F.}~\bibnamefont{Edwards}}, \bibinfo{author}{\bibfnamefont{E.}~\bibnamefont{Kendall}}, \bibinfo{author}{\bibfnamefont{S.}~\bibnamefont{Hotchkiss}}, \bibnamefont{and} \bibinfo{author}{\bibfnamefont{R.}~\bibnamefont{Easther}}, \bibinfo{journal}{JCAP} \textbf{\bibinfo{volume}{2018}}, \bibinfo{pages}{027} (\bibinfo{year}{2018}).

\bibitem[{\citenamefont{{Misner} et~al.}(1973)\citenamefont{{Misner}, {Thorne}, and {Wheeler}}}]{Misner1973}
\bibinfo{author}{\bibfnamefont{C.~W.} \bibnamefont{{Misner}}}, \bibinfo{author}{\bibfnamefont{K.~S.} \bibnamefont{{Thorne}}}, \bibnamefont{and} \bibinfo{author}{\bibfnamefont{J.~A.} \bibnamefont{{Wheeler}}}, \emph{\bibinfo{title}{{Gravitation}}} (\bibinfo{publisher}{San Francisco: W.H.~Freeman and Co.}, \bibinfo{year}{1973}).

\end{thebibliography}
\end{document}